\documentclass{aa} 
\usepackage{hyperref} 
\usepackage{acronym}
\usepackage{graphicx}
\usepackage{txfonts}
\usepackage{amsmath}
\usepackage{amssymb}
\usepackage{adjustbox}
\usepackage{tikz}
\usepackage{subcaption}
\usepackage{xcolor}
\usepackage{caption}
\usepackage{dsfont}
\usepackage{booktabs}
\usepackage{longtable}
\definecolor{mygray}{RGB}{128, 128, 128}
\definecolor{mygreen}{RGB}{0, 128, 0}
\definecolor{mylightgray}{RGB}{211,211,211}
\definecolor{myblue}{RGB}{0, 114, 178}
\usepackage[switch]{lineno}

\begin{document}
\acrodef{ACIS}{Advanced CCD Imaging Spectrometer}
\acrodef{BI}{back-illuminated}
\acrodef{CCDs} {charge-coupled devices}
\acrodef{CIAO}{Chandra Interactive Analysis of Observations}
\acrodef{CRs} {cosmic rays}
\acrodef{CXC}{Chandra X-ray Center}
\acrodef{DLR}{Deutsches Zentrum für Luft- und Raumfahrt}
\acrodef{EP}{expectation propagation}
\acrodef{FI}{front-illuminated}
\acrodef{FOV}{field of view}
\acrodef{geoVI}{geometric Variational Inference}
\acrodef{IFT}{information field theory}
\acrodef{ISM}{interstellar medium}
\acrodef{KL}{Kullback-Leibler divergence}
\acrodef{L0}{Level 0}
\acrodef{L3}{Level 3}
\acrodef{MARX}{Model of AXAF Response to X-rays}
\acrodef{MGVI}{Metric Gaussian Variational Inference}
\acrodef{MF}{multi frequency}
\acrodef{NE}{north east}
\acrodef{NW}{north west}
\acrodef{NWR}{noise-weighted residual}
\acrodef{PCA}{principal component analysis}
\acrodef{ICA}{independent component analysis}
\acrodef{PSF}{point spread function}
\acrodef{RMF}{response matrix function}
\acrodef{SE}{south east}
\acrodef{SF}{single frequency}
\acrodef{SN1a}{type 1a supernova}
\acrodef{SNR}{supernova remnant}
\acrodef{SW}{south west}
\acrodef{UBIK}{Universal Bayesian Imaging Toolkit}
\acrodef{UWRs}{uncertainty weighted residuals}
\acrodef{VI}{variational inference}

\title{First spatio-spectral Bayesian imaging of SN1006 in X-ray}
\abstract{
Supernovae are an important source of energy in the interstellar medium. Young remnants of supernovae have a peak emission in the X-ray region, making them interesting objects for X-ray observations. In particular, the supernova remnant SN1006 is of great interest due to its historical record, proximity and brightness. It has therefore been studied by several X-ray telescopes. Improving X-ray imaging of this and other remnants is important but challenging, as it often requires multiple observations with different instrument responses to image the entire object. Here, we use Chandra observations to demonstrate the capabilities of Bayesian image reconstruction using information field theory. Our objective is to reconstruct denoised, deconvolved and spatio-spectral resolved images from X-ray observations and to decompose the emission into different morphologies, namely diffuse and point-like. Further, we aim to fuse data from different detectors and pointings into a mosaic and quantify the uncertainty of our result. Utilizing prior knowledge on the spatial and spectral correlation structure of the two components, diffuse emission and point sources, the presented method allows the effective decomposition of the signal into these. In order to accelerate the imaging process, we introduce a multi-step approach, in which the spatial reconstruction obtained for a single energy range is used to derive an informed starting point for the full spatio-spectral reconstruction. The method is applied to 11 Chandra observations of SN1006 from 2008 and 2012, providing a detailed, denoised and decomposed view of the remnant. In particular, the separated view of the diffuse emission should provide new insights into its complex small-scale structures in the center of the remnant and at the shock front profiles. For example, our analysis shows sharp X-ray flux increases by up to two orders of magnitude at the shock fronts of SN1006.}
\author{M. Westerkamp
          \inst{1, 2}          \and
        V. Eberle
          \inst{1, 2}
          \and
        M. Guardiani
          \inst{1, 2}
          \and 
        P. Frank
          \inst{1}
          \and
        L. Scheel-Platz
          \inst{1, 2, 3, 4, 5}
        \and
        P. Arras
          \inst{1}
          \and
        J. Knollmüller
          \inst{6, 7}
          \and
        J. Stadler
          \inst{1, 6}
          \and
        T. En{\ss}lin
          \inst{1, 2, 6}
          }
\institute{Max Planck Institute for Astrophysics,
           Karl-Schwarzschild-Str. 1, 85748 Garching, Germany\\
           \and
		   Faculty of Physics, Ludwig-Maximilians-Universitaet Muenchen 				   (LMU), Geschwister-Scholl-Platz 1, 80539 Munich, Germany \\
		   \and
		   Institute of Biological and Medical Imaging,   Helmholtz Zentrum   		   Muenchen,  Ingolstaedter Landstr. 1, 85764 Neuherberg, Germany \\
		   \and 
		   Institute of Computational Biology,  Helmholtz Zentrum Muenchen,  		   Ingolstaedter Landstr. 1, 85764 Neuherberg, Germany \\
		   \and		   
		   Technical University of Munich,   School of Medicine, 				 		   Einsteinstr. 25, 81675 Munich, Germany \\
		   \and 
		   Excellence Cluster ORIGINS, Boltzmannstr. 2, 85748 Garching, 				   Germany \\
		   \and
		   Technical University of Munich, TUM School of Natural Sciences,
    	   Boltzmannstr. 2, 85748 Garching, Germany}
\date{Received \today ;}
\maketitle
\section{Introduction}
\nolinenumbers
In the year 1006, observers on Earth were able to see the light of a bright "new star", which eventually faded after a few months. This observation is now attributed to a \ac{SN1a} event that produced its remnant, known as SN1006 or \ac{SNR} G327.6+14.6. It is the brightest stellar event ever recorded. Its historical record \citep{https://doi.org/10.1111/j.1468-4004.2010.51527.x} is one of the reasons why this remnant was an interesting target for several observational campaigns. SN1006 is notable for being a relatively unobscured \ac{SNR} \citep{Katsuda_2013}, which is large in angular size due to its proximity to Earth \citep{Winkler_2003}. All these points made SN1006 a good object to study \ac{SN1a} events and led to an impressive research history. \\
In particular, X-ray observations of the remnant have provided important information about the dynamics and energies of the supernova explosion and the surrounding interstellar medium. When a supernova explodes, it creates a rapidly expanding shell of ejected material that compresses and aggregates up the surrounding \ac{ISM}. The collision between the expanding shell and the \ac{ISM} creates a shock wave that propagates into the \ac{ISM} and heats it up so that it emits thermal and non-thermal X-rays \citep{seward_charles_2010}. In young \ac{SNR}s, both thermal and non-thermal emission have a peak in the 0.5-10 keV energy range \citep{10.1093/mnras/stv1882}, making current X-ray telescopes perfect for studying these objects. An important observation was made by \citet{koyama_evidence_1995}, who detected synchrotron X-ray emission in the envelope of SN1006, supporting the theoretical expectation that the shock wave of \ac{SNR}s accelerates particles to extremely high energies. This is believed to be a major production process of \ac{CRs}. Accordingly, \ac{SNR}s are one important source of energy for the \ac{ISM} via cosmic rays. This observation led to many subsequent spectral \citep{Helder_2012,doi:10.1146/annurev.astro.46.060407.145237} and spatio-spectral analyses of SN1006 \citep{2003ApJ...589..827B, winkler2014high, 10.1093/mnras/stv1882} to study the spatially varying X-ray production processes in SN1006. In addition, supernovae are known to produce heavier elements from lighter ones during the explosion, which are ejected into the \ac{ISM} and enable the formation of new stars and planetary systems, making them very important for the Galactic metabolism. \citet{winkler2014high} and \citet{10.1093/mnras/stv1882} have studied the spatial distribution of elements in the remnant. Long-term observations of SN1006 allowed \citet{winkler2014high} and \citet{Katsuda_2013} to study its proper motion and thereby gave insight into the dynamics of the explosion, the evolution of the remnant, and its interaction with the interstellar medium. Despite the extensive previous studies of SN1006 and other \ac{SNR}s, there are still a number of aspects that are not well understood. Among them are the details of particle acceleration at shock fronts \citep{Vink_2011}.\\
In recent years, there have been significant advances in X-ray astronomy aimed at studying such high-energy phenomena in the universe. These advances have been driven in large part by the development of new X-ray satellite missions such as Chandra, XMM-Newton and Suzaku, which have provided unprecedented spatial and spectral resolution. However, any technological advance in space-based astronomical instruments must be accompanied by advances in imaging methodology in order to exploit the full potential of these instruments. Here, we focus on the development of such an imaging method, capable of denoising, deconvolving and decomposing the data, and apply it to Chandra observations of SN1006 - the highest resolution data of this \ac{SNR} to date.  The aim is to obtain a more detailed view of the small-scale structures of the remnant, and thus to allow a more detailed study of the open questions in the field of supernovae and their remnants as well as to challenge and benchmark the imaging method. \\
To obtain an accurate and meaningful reconstruction of the true flux from the given X-ray data, there are a number of challenges that need to be overcome. X-ray telescopes like Chandra record the data from these high-energy phenomena as photon count events accompanied by information about the photon's arrival direction, time, and energy. In this work, the events are categorized into spatial and energy bins, which yields independent Poisson statistics for each pixel. In particular, X-ray observations often have low count rates, which poses a challenge because of the resulting poor signal-to-noise ratio. Accordingly, 
a major task in X-ray imaging is the denoising of the corresponding data. In addition, there is an instrument specific response to the observed X-ray flux, which complicates the relation between the sky and data and includes in particular the exposure and the \ac{PSF}. A coherent representation and application of the response is a complex problem, as the instrumental properties of X-ray instruments tend to change with off-axis angle, energy and time. Ultimately, one of the goals of X-ray imaging is to discriminate between noise, background, and extended and point sources. So far, most imaging techniques are designed to extract either the point sources or the diffuse flux, but lack the ability to reconstruct both simultaneously. Especially for the study of extended sources like SN1006, this separation of components is essential to study the spectra and thus the emission properties of the remnant at each location.\\
The here presented study aims to address these challenges in X-ray imaging. In particular, we use \ac{IFT} \citep{ensslin2019information} as a versatile mathematical framework for reconstructing the signal from large and noisy data sets. \ac{IFT} combines information theory, statistical physics and probability theory. Together with the numerical \ac{IFT} algorithms implemented in the software package NIFTy \citep{arras2019nifty5}, it provides an excellent tool for denoising, deconvolving and decomposing the image, as already demonstrated for Poisson data \citep{Selig_2015, D4PO, platz2022multicomponent}. The basis of \ac{IFT} is Bayes theorem applied to the problem of reconstructing fields. In our case, the sky photon flux is regarded to be a field, which we subsequently refer to as the signal field. It is inferred given prior knowledge on its configuration and the measurement data, which is interpreted in the light of a model for the measurement response. The instrument description including its noise statistics determines the so called likelihood, in other words the probability to observe specific data given a sky flux configuration. By combining the prior and the likelihood into the posterior distribution, we obtain not only an estimate of the actual sky photon flux as its posterior mean, but also an estimate of the uncertainty via the posterior variance. \\
During inference, the prior model guides the separation of the signal into different components, such as point-like and diffuse structures. Thus, we need to carefully encode our knowledge of the different components into our prior model, to give the inference the chance to discriminate their contributions to the observed photon counts. To this end, we model the signal field as a superposition of different physical fluxes: the emission from point-like and extended sources. Assigning a different correlation structure to the diffuse emission from extended sources, which is assumed to be spatially correlated, and the point sources, which are assumed to be spatially uncorrelated, makes it possible to distinguish between these components. A spatio-spectral prior allows the reconstruction of the emissivity as a function of energy and spatial position. Further knowledge about the different spectra of the components improves their separation.\\
The instrumental description encoded in the likelihood drives the deconvolution of the data from the \ac{PSF}, the image denoising and the exposure correction. Specifically for Chandra, there are two different ACIS X-ray imagers, ACIS-I and ACIS-S. The majority of the chips in ACIS-I and ACIS-S are front illuminated. However, ACIS-S also contains two chips that are back-illuminated leading to a significant number of non-astronomical photon events in these regions. To account for the latter, we added a further model component of a non-astronomical, spatially varying but temporally constant background that is present in regions of the back-illuminated chips. An additional challenge is the fact that Chandra's \ac{FOV} is small compared with the extent of SN1006. It is therefore not possible to capture the entire remnant in a single ACIS-I or ACIS-S image. Instead, mosaicing is required \citep{winkler2014high}. This mosaicing can be effectively implemented, even for varying instrument responses, by combining the corresponding likelihoods. \\
Overall, the spatio-spectral inference of the sky flux is associated with a significantly higher computational complexity than an inference that considers only the spatial direction. Therefore, we introduce a multi-step model, which considers two different priors, a purely spatial one and a spatio-spectral one. First, we perform a spatial reconstruction using the spatial prior. The result of this spatial reconstruction is mapped onto the entire spatio-spectral sky. The mapped sky with multiple energy bins added is used as the initial guess for the subsequent spatio-spectral reconstruction. This allows us to perform parts of the reconstruction and especially of the component separation  in a smaller parameter space.\\
This multi-step model, which we will call the transition model, and the reconstruction results on SN1006 are presented and discussed in the following. The structure of the remaining sections is as follows. In Sect. \ref{sec:related_work} we present current methods used in X-ray imaging and their application results on SN1006 data so far. We also review the state of the art for photon count data in the field of \ac{IFT}. An introduction to the imaging of photon data with \ac{IFT} is given in Sect. \ref{sec:IFT}. The explicit structure of the algorithm and in particular of the transition model is given in Sect. \ref{sec:InferenceAlgorithm}. Sect. \ref{sec:Priors} focuses on the corresponding prior description and Sect. \ref{sec:InstrumentResponse} explains the instrument model and the Chandra observations of SN1006. In Sect. \ref{sec:mock} we present a reconstruction from synthetic data to validate the method, before finally presenting and discussing the reconstruction results on SN1006 in Sect. \ref{sec:results}. The conclusion and outlook for future research is part of Sect. \ref{sec:conclusion}.
\section{Related work}
\label{sec:related_work}
This section is devoted to a review of previous studies and the state of the art in X-ray imaging. Previous investigations in high-energy astrophysics, with a focus on X-ray studies, are highlighted in three parts - A discussion of previous and current X-ray imaging techniques, the explanation of the results of these techniques applied to SN1006, an introduction to previous imaging techniques with \ac{IFT} on which the reconstruction presented here is based.
\subsection{State of the art: X-ray imaging}
\label{sec:stateoftheartxray}
The study of X-ray phenomena in the universe began in the 1960s and is a relatively new field of astrophysics, due to the inability of ground-based telescopes to observe X-rays from astronomical sources. However, there have been many technical developments since then, which are discussed in more detail in \citet{seward_charles_2010}. Here, we will focus on the imaging techniques that have been developed and are in use, with a non-exclusive focus on the Chandra X-ray Observatory. A more comprehensive summary of recent developments in X-ray analysis for the X-ray telescopes XMM-Newton, Suzaku and Chandra can be found in \citet{seward_charles_2010}. \\
Among others, \citet{seward_charles_2010} give an insight into the steps and techniques in the widely used Chandra data processing pipeline\footnote{https://cxc.harvard.edu/ciao/dictionary/sdp.html}. The corresponding methods and further data imaging and response tools are implemented in the software tool \ac{CIAO}. \citep{CIAO}, developed by the \ac{CXC}.\\
Overall, there are some standards for extended sources such as SN1006 that are used in recent publications. One of these is the reduction of background from the data, which can obscure the signal from the source of interest. A disadvantage of this approach is that the subtraction of the background comes at the cost of eliminating real X-ray events. Another tool, used particularly for extended sources, is mosaicing. This allows the analysis of sources, which have a larger extent than Chandra's \ac{FOV}. In \ac{CIAO} mosaicing is implemented by transforming the raw count images, the effective area and the background maps into a single coordinate system. Reconstructing an image from these mosaics has its difficulties, as there are often several \ac{PSF}s and \ac{RMF}s for one source. So far, this problem has been overcome by calculating and using the weighted average of the \ac{PSF} and \ac{RMF} for the data patches, as suggested by \citet{Broos_2010}. \\
One of the final steps, which depends on the object of interest, is source detection and extraction. The aim is to separate the X-ray source of interest from the background. For this purpose, three well-known algorithms, the sliding cell algorithm \citep{2001ASPC..238..443C}, the wavelet detection algorithm \citep{wavelets2008} and the Voronoi tessellation and percolation algorithm \citep{PhysRevE.47.704}, are implemented in \ac{CIAO}. The sliding cell algorithm, already used for Einstein and ROSAT, searches for sources by summing the counts in a square cell that slides over the image. For comparison, the counts in a cell assigned to the background are taken. From the ratio of the counts in the cell to the counts in the background the cell might be assigned to a source. Wavelet detection, on the other hand, decomposes the signal into a series of wavelets. By analyzing the coefficients of the wavelets, patterns of different scales can be detected in the data. Finally, data cleaning and source extraction techniques differ for point sources and diffuse emission. This involves additional work as the pipeline needs to be run several times to fully extract point source and diffuse emission information. \\
There have been other approaches to source decomposition that fall into the category of blind source separation. In general, the goal of blind source separation is to automatically decompose observations into features maximizing their statistical separation.
In \citet{2005ApJ...634..376W} a \ac{PCA} approach is presented to determine the location of the contact discontinuity and the shock wave, and thus find evidence for cosmic ray acceleration in the \ac{SNR} Tycho.  In particular, sparse blind source separation aims to compress the signal and thereby extract its essence. One application of sparse blind source separation to Chandra data was recently presented in \citet{Picquenot_2019}. Moreover, generalized morphological component analysis \citet{Bobin2016} was performed to X-ray data of Cassiopeia A by \citet{Bobin2020SparseBF} to decompose the spectrum into its components like thermal and non-thermal emission. Here, the generalized morphological component analysis models the source as a linear combination of a fixed number of morphological components and solves the according blind source separation problem, while putting sparsity constraints on the morphological components. \\
In addition, Bayesian and machine learning approaches have been applied for source separation, model comparison, or point source characterization. \citet{Guglielmetti_2009} analyzed Bayesian techniques for the joint estimation of sources and background, and \citet{1988ESOC...28..177C} implemented a maximum likelihood algorithm for the calculation of certain parameters of the detected sources, which is also used for XXM-Newton. In \citet{Ellien_2023} different components of the spectrum are modeled for Chandra data of five thin bands around Tycho and different one-, two-, and three-component models are analyzed by  Bayesian model comparison. Recently, a machine learning approach has been published by \citet{10.1093/mnras/stad414}, which is intended to work as an automated source classifier. The approach is based on supervised learning and allows point sources to be assigned to specific classes. 
\subsection{Previous studies of SN1006 in the X-ray range}
The supernova remnant SN1006 has an exciting scientific record. As mentioned above, the remnant is of great scientific interest in the study of Type 1a supernovae and their remnants for many reasons - its proximity, its low obscuration, and its large size. In particular, X-ray observations of the remnant provide an opportunity to study its evolution. Accordingly, there have been intensive studies of SN1006 in this energy range, starting with observations by ROSAT \citep{10.1093/mnras/278.3.749} and ASCA \citep{koyama_evidence_1995}. The ASCA data on SN1006 were analyzed by \citet{koyama_evidence_1995} and \citet{Dyer2003SeparatingTA}, which led to the confirmation of theoretical predictions that cosmic rays are accelerated at the shock fronts of the remnant. It was also discovered that there are several processes in the supernova remnant that are responsible for the X-ray emission. In fact, it was found that the \ac{NE} and \ac{SW} of SN1006 are dominated by non-thermal, synchrotron emission, while the \ac{NW} and \ac{SE} edges are less distinct and are attributed to thermal emission. Accordingly, \citet{Dyer2003SeparatingTA} analyzed non-thermal and thermal models on the ASCA data. \\
The new technologies of the X-ray telescopes XMM-Newton and Chandra have led to an unprecedented resolution of X-ray sources and thus to better data of SN1006. \citet{2003ApJ...589..827B} published the first spatio-spectral study of Chandra ACIS-S data from the \ac{NE} shell of SN1006, followed by ACIS-I mosaic data from the analysis of \citet{Cassam_Chena__2008}. Here, we want to highlight the publication of \citet{winkler2014high}, as their reconstructed image should be the main point of comparison for ours. In \citet{winkler2014high} the standard Chandra pipeline as described in Sect. \ref{sec:stateoftheartxray} was used and point sources were extracted using the wavelet detection algorithm. To use multiple data patches, the observations were merged using \ac{CIAO}. A more recent study of SN1006 was carried out by \citet{10.1093/mnras/stv1882} on XMM-Newton data using the SAS software. The data were preprocessed in a similar way to the Chandra pipeline and wavelet detection was used. However, point source detection was only possible at high energies because of the risk of misidentifying small-scale structures in the low energy regime as point sources.

\subsection{Previous work on high energy count data with \ac{IFT}}
High energy astronomical data, including X-ray and gamma-ray data, are recorded in photon counts. So far there have been no applications of \ac{IFT} to Chandra X-ray data, but there have been studies on gamma-ray data and methodological research on the reconstruction and component separation of such count data. First, the algorithm D3PO by \citet{Selig_2015} based on \ac{IFT} implemented the denoising, deconvolution and decomposition of count data. Building on this, D4PO by \citet{D4PO} allows D3PO to work on fields that have spectral and temporal coordinates in addition to spatial coordinates. Finally, \citet{platz2022multicomponent} built a model of the gamma-ray sky and applied a variant of D4PO in a spatio-spectral setting. \\
In this work, we adapt a similar model as presented in \citet{platz2022multicomponent} to describe the X-ray sky. As such, this is the first application of \ac{IFT} imaging to X-ray data. Further, we introduce a method to fuse several data sets with different detector characteristics, pointing directions and noise levels into a mosaic. We demonstrate how the imaging can be accelerated and improved by a multi-step model, which is presented in Sect. \ref{sec:InferenceAlgorithm}.

\section{Image reconstruction with \ac{IFT}}
\label{sec:IFT}
In X-ray imaging we deal with finite, incomplete and noisy data. We use \ac{IFT} \citep{ensslin2019information}, an information theory for fields, to infer the X-ray sky as a continuous field from this finite data $d$. In general, a physical field, $s: \Omega \to \mathbb{R}$, assigns a value to each point in the space $\Omega$, which describes a continuous physical quantity such as temperature, pressure, intensity etc. or in our case X-ray flux. Given the data $d$, we obtain constraints on the field of interest, which we call the signal field. 
Since the data provides only a finite number of constraints on the signal field, there could have been an infinite number of signals that have produced the data, even if we completely neglect noise. For this reason, prior assumptions about the field are needed to sufficiently constrain the signal field $s$. Given the likelihood $\mathcal{P}(d|s)$, which describes the measurement, and a statistical description of the prior, $\mathcal{P}(s)$, the posterior probability of the signal given the data can be calculated via Bayes theorem,
\begin{align}
\label{eq:Bayes}
\mathcal{P}(s|d)=\frac{\mathcal{P}(d|s)\mathcal{P}(s)}{\mathcal{P}(d)}.
\end{align} 
In \ac{IFT} we aim to inspect this posterior probability, as it allows us to draw posterior samples and thereby calculate any important posterior quantity such as the posterior mean,
\begin{align}
\label{eq:PosteriorMean}
m = \langle s \rangle_{(s|d)} \equiv \int \mathcal{D}s ~s~ \mathcal{P}(s|d),
\end{align}
where $\int \mathcal{D}s$ denotes the path integral over all possible field configurations and a measure of uncertainty via the covariance of the posterior probability,
\begin{align}
\label{eq:PosteriorCovariance}
D = \langle (m-s)(m-s)^\dagger \rangle_{(s|d)}.
\end{align}
The expectation value over the posterior probability is denoted by $\langle\rangle_{(s|d)}$ and $\dagger$ gives the adjoint of the corresponding field. Therefore, the statistical treatment of the fields of interest in \ac{IFT} creates an important advantage, as we can not only present a point estimate of the field, but also quantify its reliability at each position.\\
A more detailed description of the likelihood and prior model is given in Sect. \ref{sec:InferenceAlgorithm}. Here, we describe image reconstruction with \ac{IFT} given a general measurement equation. Accordingly, we consider a measurement as a function $f$ that maps a field from its continuous space to a discrete data space. This function is determined by the response, $R(s)$, of the instrument and some statistical noise, $n$, in the measurement,
\begin{align}
\label{eq:GeneralMeasurementEquation}
d=f(R(s),n).
\end{align} 
Given this generic measurement equation we can calculate the likelihood by marginalizing over the measurement noise,
\begin{align}
\label{eq:GeneralLikelihoodEquation}
\mathcal{P}(d|s) &= \int dn ~ \mathcal{P}(d, n|s)  = \int dn ~ \mathcal{P}(d|n, s)~ \mathcal{P}(n|s) \\ 
&= \int dn ~ \delta(d-f(R(s),n))~\mathcal{P}(n|s) \\
&= \mathcal{P}\left(f^{-1}(R(s),~d)|s\right) ~\bigg \vert \frac{\partial f(R(s),n)}{\partial n}  \bigg \vert^{-1}
\end{align}
where $f^{-1}$, is the inverse of the measurement function with respect to the second argument $n$ and $\vert \partial f / \partial n \vert$ is the functional determinant.
When combining the likelihood with a prior distribution to obtain the posterior, the main difficulty lies in normalizing the posterior, i.e. in computing the evidence $\mathcal{P}(d)= \int \mathcal{D}s ~\mathcal{P}(d|s) \mathcal{P}(d)$. To circumvent the problem of analytically intractable normalization, we approximate the posterior via \ac{VI}, where a possibly complex posterior distribution $\mathcal{P}(s|d)$ is approximated by a simpler one, $\mathcal{Q}(s| d)$. Mathematically, the \ac{KL} \citep{10.1214/aoms/1177729694} is the measure that needs to be optimized to find the optimal approximation,
\begin{align}
\label{eq:KL}
\mathcal{D}_\text{KL}(\mathcal{Q}(s|d)\vert \mathcal{P}(s|d)) = \int \mathcal{D}s ~\mathcal{Q}(s|d) \ln\bigg(\frac{\mathcal{Q}(s|d)}{\mathcal{P}(s|d)}\bigg).
\end{align}
Here, we use the \ac{VI} version of the \ac{KL} divergence, which, if minimized for the parameters of $\mathcal{Q}(s|d)$, ensures that in the approximation the least amount of spurious information is introduced. The \ac{EP} version of the \ac{KL} divergence, $\mathcal{D}_\text{KL}(\mathcal{P}(s|d)\vert \mathcal{Q}(s|d))$, would be more conservative, as it would just ensure that a minimum of information is lost, but none is introduced. However, \ac{EP} requires integration over the intractable posterior $\mathcal{P}(s|d)$, while \ac{VI} only requires integration over a conveniently chosen function $\mathcal{Q}(s|d)$ (e.g. a Gaussian), and therefore is feasible. As a consequence of this, uncertainty estimates obtained from the \ac{VI} approximation are known to be a bit too optimistic, which should be kept in mind. However, those are nevertheless well informative about the structure of the uncertainties. For further details we refer the reader to \citep{Frank_2021}.
\section{Algorithm overview of Bayesian inference of the X-ray sky}
\label{sec:InferenceAlgorithm}
\subsection{Structure of the reconstruction algorithm}
\label{sec:StructureOfAlgorithm}
The measure of interest in our reconstruction of SN1006 is the sky flux $s$ as a function of space and energy. In other words, the signal field $s$ lives on a space consisting of a two-dimensional position space and a one-dimensional log-energy space, denoted by $z=(\vec{x}, y) ~ \in \Omega=\mathbb{R}^2\times\mathbb{R}$, where $y=\log(E/E_0)$ and $E_0$ is the reference energy. In order to guide the inference in the latent space and to reduce computational complexity, we introduce a multi-step model, which we call the transition model. The transition model divides the actual reconstruction into three parts, with three different inference problems, which are solved by \ac{VI}. First, we aim to reconstruct the sky at a single energy level. Here, we perform a purely spatial reconstruction of the signal of interest. This part of the reconstruction is called the \ac{SF} reconstruction. Its results are used to determine an informed starting position for the spatio-spectral reconstruction, subsequently called the \ac{MF} reconstruction. The standard reconstruction algorithm for the \ac{SF} and \ac{MF} model are further described in Sect. \ref{sec:StandardInference}. We model the mapping from the \ac{SF} image to the \ac{MF} image space as an inference problem, whose solution constitutes step two, which is introduced in Sect. \ref{sec:Transition}. In the third step, we solve the \ac{MF} reconstruction using the starting point provided by step two. A similar model was previously used by \citet{Arras_2022} to move from a spatial domain to a spatio-temporal domain. \\
Using the transition model, we solve significant parts of the reconstruction problem, including the separation of point source and diffuse emission, in the \ac{SF} setting, which has less model and computational complexity. Table \ref{tab:parameter_table} shows the number of hyper-parameters for each model, \ac{SF} and \ac{MF}, and its sub-components, reflecting the model complexity, while Table \ref{tab:latent_parameter_table} presents the number of latent parameters in the model as a measure for the computational complexity. Fig. \ref{fig:comptime} ans \ref{fig:compacc} in the appendix show a quantitative comparison regarding computational complexity and reconstruction error for the transition model presented here versus a pure MF reconstruction. A schematic overview of the described reconstruction algorithm can be seen in the diagram in Fig. \ref{fig:transition model}. The \ac{MF} and \ac{SF} prior models themselves are discussed in Sect. \ref{sec:Priors}.
\subsection{Variational inference and generative models}
\label{sec:StandardInference}
As mentioned in Sect. \ref{sec:IFT} we approximate the posterior given information on the prior model, which is subject of Sect. \ref{sec:Priors}, 
and the likelihood description, which is elaborated in Sect. \ref{sec:InstrumentResponse}, via \ac{VI}. Two approaches to \ac{VI} of posteriors within the current NIFTy package are \ac{MGVI} \citep{knollmueller2020metric} and \ac{geoVI} \citep{Frank_2021}. They are designed to approximate high-dimensional and complex posterior probability distributions via optimization of the cross entropy term of the \ac{KL} in Eq.~\eqref{eq:KL}. Both approaches perform the \ac{KL} optimization in a coordinate space of the problem, in which the prior is a standard Gaussian. In particular, the signal field is described by a generative model $s = s(\xi)$ given a set of latent parameters $\xi$ with a standard Gaussian prior $\mathcal{P}(x_i) = \mathcal{G}(\xi_i, \mathds{1})$. The generative model encodes all prior knowledge on the corresponding field. To this end, the likelihood is formulated as a function of the latent parameters, $\mathcal{P}(d|\xi)$ and the posterior $\mathcal{P}(\xi|d)$ can be inferred via \ac{VI}. In this work, \ac{geoVI} is used, which optimizes the \ac{KL} for the parameters of a non-linear coordinate transformation in which the posterior becomes an approximate standardized Gaussian. Thereby, \ac{geoVI} allows for the representation of non-Gaussian signal posteriors. The detailed implementation can be found in \citep{Frank_2021}.
In any case, we need to define generative prior models for both, the \ac{SF} and the \ac{MF} model, given the corresponding latent parameters $s_m = s_m(\xi_m)$, $m \in \{\text{SF}, \text{MF}\}$. The detailed explanation of these models is part of the prior description in Sect. \ref{sec:Priors}. The according posterior approximations for each model $m$ are denoted by $\mathcal{Q}_m$.
\subsection{The transition model}
\label{sec:Transition}
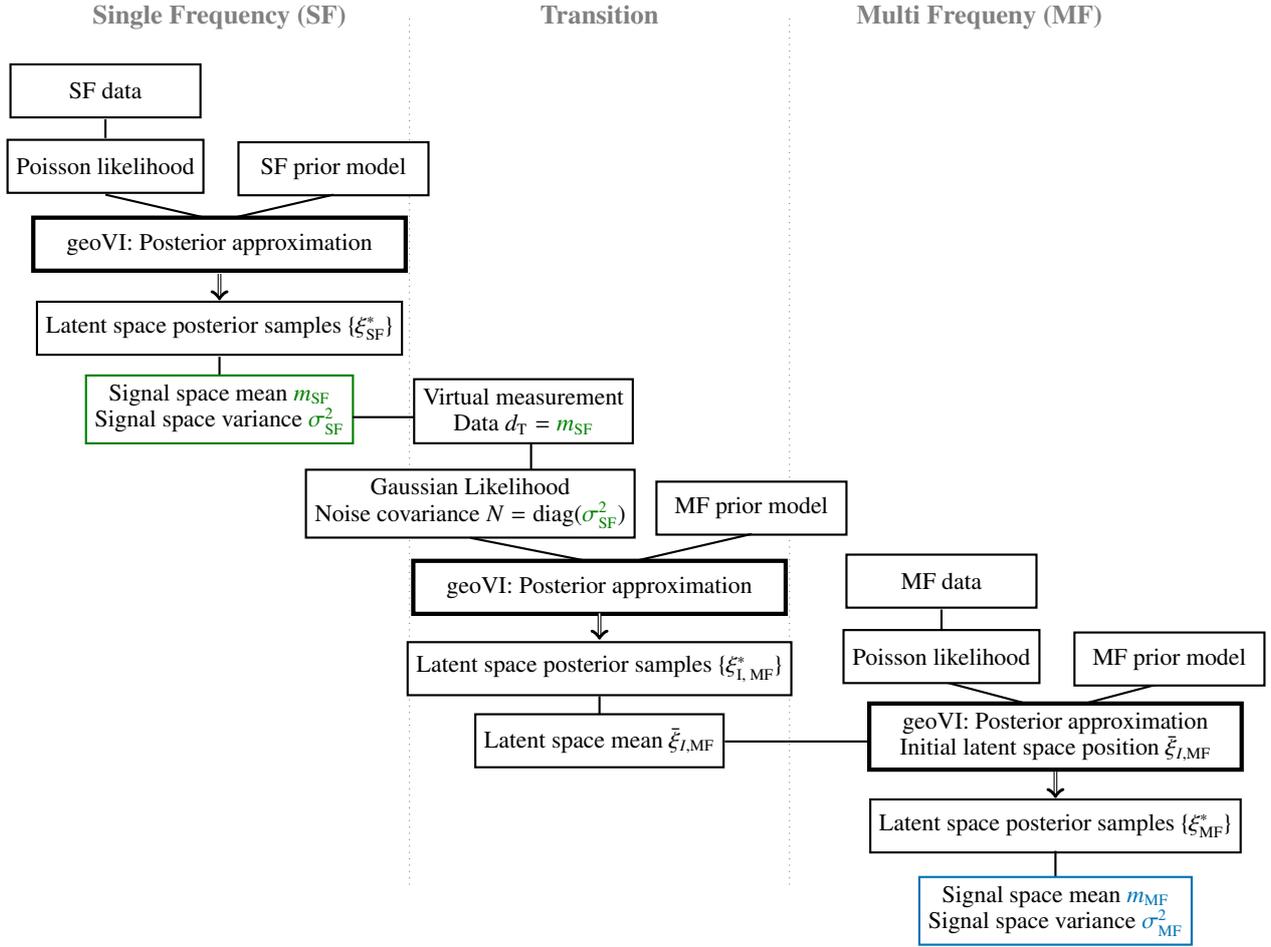
\begin{figure*}
\centering
\begin{tikzpicture}[
  every node/.style={draw, anchor=west, thick},
  arrow/.style={thick},
  normalcenter/.style={draw, anchor=mid, minimum width=49mm, align=center, minimum height=2em, fill=white, font=\small},
  smallcenter/.style={draw, anchor=mid, minimum width=25mm, align=center, minimum height=2em, fill=white, font=\small},
  ]
  
\draw[mygray, dotted] (5,8.5) -- (5.0,20); 
\draw[mygray, dotted] (10,8.5) -- (10.0,20); 

\node[mygray, draw=none, anchor=mid, align=center, font=\bfseries] at (2.5, 20){Single Frequency (SF)};
\node[mygray, draw=none, anchor=mid, align=center, font=\bfseries] at (7.5, 20){Transition};
\node[mygray, draw=none, anchor=mid, align=center, font=\bfseries] at (12.5, 20){Multi Frequeny (MF)};

\node [smallcenter] (sky_1) at (1.0, 19.0) {SF data};
\draw[arrow] (1.0, 18.65) -- (1.0, 18.4);
\node [smallcenter] (sky_1) at (1.0, 18.0) {Poisson likelihood};
\draw[arrow] (1.0, 17.65) -- (2.5, 17.3);
\node [smallcenter] (sky_1) at (4.0, 18.0) {SF prior model};
\draw[arrow] (4.0, 17.65) -- (2.5, 17.3);
\node [normalcenter, ultra thick] (sky_1) at (2.5, 17.0) {geoVI: Posterior approximation};
\draw[->, double] (2.5, 16.6) -- (2.5, 16.25);
\node [smallcenter] (m_sf) at (2.5, 15.9) {Latent space posterior samples $\{\xi_{\text{SF}}^*\}$};
\draw[arrow] (2.5, 15.5) -- (2.5, 15.2);
\node [draw=mygreen, smallcenter] (m_sf) at (2.5, 14.65) {Signal space mean $\textcolor{mygreen}{m_\text{SF}}$\\
Signal space variance $\textcolor{mygreen}{\sigma_\text{SF}^2}$};
\draw[arrow] (4.25, 14.7) -- (5.1, 14.7);
\node [smallcenter] (m_sf) at (6.5, 14.6) {Virtual measurement\\
Data $d_\text{T} = \textcolor{mygreen}{m_\text{SF}}$};
\draw[arrow] (6.6, 14.35) -- (6.6, 14.0);
\node [smallcenter] (m_sf) at (5.8, 13.4) {Gaussian Likelihood\\
Noise covariance $N = \text{diag}(\textcolor{mygreen}{\sigma_\text{SF}^2})$};
\draw[arrow] (5.8, 13.1) -- (7.3, 12.8);
\node [smallcenter] (m_sf) at (9.5, 13.5) {MF prior model};
\draw[arrow] (9.5, 13.15) -- (8, 12.8);
\node [normalcenter, ultra thick] (sky_1) at (7.5, 12.45) {geoVI: Posterior approximation};
\draw[->, double] (7.5, 12.1) -- (7.5, 11.75);
\node [smallcenter] (xi_sf) at (7.5, 11.4) {Latent space posterior samples $\{\xi_{\text{I, MF}}^*\}$};
\draw[arrow] (7.5, 11.0) -- (7.5, 10.75);
\node [smallcenter] (xi_sf) at (7.5, 10.4) {Latent space mean $\bar{\xi}_{I,\text{MF}}$};
\draw[arrow] (9.15, 10.4) -- (11.1, 10.4);
\node [smallcenter] (sky_2) at (12.0, 12.5) {MF data};
\draw[arrow] (12.0, 12.15) -- (12.0, 11.85);
\draw[arrow] (12.0, 11.2) -- (13.5, 10.8);
\draw[arrow] (15.0, 11.2) -- (13.5, 10.8);
\node [smallcenter] (sky_2) at (12.0, 11.5) {Poisson likelihood};
\node [smallcenter] (sky_1) at (15.0, 11.5) {MF prior model};
\node [normalcenter, ultra thick] (sky_1) at (13.5, 10.3) {geoVI: Posterior approximation\\
Initial latent space position $\bar{\xi}_{I,\text{MF}}$};
\draw[->, double] (13.5, 10.0) -- (13.5, 9.65);
\node [smallcenter] (m_sf) at (13.5, 9.3) {Latent space posterior samples $\{\xi_{\text{MF}}^*\}$};
\draw[arrow] (13.5, 8.95) -- (13.5, 8.6);
\node [draw=myblue, smallcenter] (m_sf) at (13.5, 8.0) {Signal space mean $\textcolor{myblue}{m_\text{MF}}$\\
Signal space variance $\textcolor{myblue}{\sigma_\text{MF}^2}$};
\end{tikzpicture}
\caption{Structure of the transition model given the generative prior models for the \ac{SF} reconstruction $s_{\text{SF}} = s_{\text{SF}}(\xi_\text{SF})$ and for the \ac{MF} reconstruction $s_{\text{MF}} = s_{\text{MF}}(\xi_\text{MF})$, which transform the according latent parameters $\xi_\text{SF}$ and $\xi_\text{MF}$ from the latent space into the signal space. Here, $\xi_{\text{I, MF}}$ denotes the initial position of the \ac{MF} reconstruction in latent space.}
\label{fig:transition model}
\end{figure*}

The indirect encoding of fields in generative models complicates the transition from one model (e.g. the \ac{SF} model) to another (e.g. the \ac{MF} model) as the corresponding generative function is in general not invertible, in other words its inverse is not unique. 
Thus, our objective is to determine a mapping function $T$ that plausibly maps the parameters of the \ac{SF} model, $\xi_{\text{SF}}$, to their corresponding \ac{MF} parameters $\xi_{\text{MF}}$, 
\begin{align}
\label{eq:ParameterMapping}
T:\xi_{\text{SF}} \to \xi_{\text{MF}}.
\end{align}
As the transition model is intended to be flexible and adaptable to a range of initial and final models, we implement it as an inference problem. Given the posterior signal space mean $m_{\text{SF}} = \langle s_{\text{SF}}(\xi) \rangle_{\mathcal{Q}_\text{SF}}$ and signal space variance $\sigma^2_{\text{SF}} = \langle (s_\text{SF}(\xi_\text{SF})- m_{\text{SF}})^2 \rangle_{\mathcal{Q}_\text{SF}}$ of the \ac{SF} reconstruction, we infer the corresponding latent space parameters of the \ac{MF} model, which we take as the starting point $\xi_{I,\text{MF}}$ for the \ac{MF} reconstruction.  The according virtual measurement equation is,
\begin{align}
\label{eq:VirtualMeasurement}
d_\text{T} = m_\text{SF} = \mathcal{R} s_\text{MF}(\xi_{I,\text{MF}})+n, ~ n \curvearrowleft \mathcal{G}(n,N),
\end{align}
where $N=\text{diag}(\sigma^2_\text{SF})$. The transition response $\mathcal{R}$ is a linear operator that can be chosen adaptively according to the problem under consideration. In the present analysis, $\mathcal{R}$ is an operator that extracts the highest energy bin from the spatio-spectral field $s_\text{MF}$, generating a two-dimensional field, $s_\text{SF}$. The likelihood of the mapping inference problem is given by the linear measurement equation in Eq.~\eqref{eq:VirtualMeasurement} and the likelihood derivation in Sect. \ref{sec:IFT} is described by a Gaussian,
\begin{align}
\mathcal{P}(d_\text{T}|\xi_{I,\text{MF}}, \sigma^2_\text{SF}) = \mathcal{G}(d_\text{T}-\mathcal{R} s_\text{MF}(\xi_{I,\text{MF}}), \text{diag}(\sigma^2_\text{SF})).
\end{align}
The posterior for the initial latent parameters in the \ac{MF} model $\mathcal{P}(\xi_{I,\text{MF}}| d_\text{T}, \sigma^2_\text{SF})$ is approximated by $\mathcal{Q}_\text{T}(\xi_{I,\text{MF}}| d_\text{T}, \sigma^2_\text{SF})$ with geoVI as described in Sect. \ref{sec:StandardInference}. We choose the posterior mean of the transition $\langle \xi_{I,\text{MF}} \rangle_{\mathcal{Q}_\text{T}}$ as the initial position for the subsequent \ac{MF} reconstruction.
This results in an overall algorithm that starts with a high energy slice and uses this reconstruction as a starting point for the subsequent spatio-spectral reconstruction. The flow of reconstructions in this approach is illustrated in Fig. \ref{fig:transition model}. The decision to start with the high energy slice was deliberate, as this particular energy range has a more consistent effective area for Chandra. Since the transition result is used only as an initial guess, we assume a consideration of a diagonal transition noise covariance $N$ to be sufficient. The possibly underestimated noise level is corrected in the subsequent spatio-spectral reconstruction steps.\\

\section{Prior models for the X-ray sky}
\label{sec:Priors}
\subsection{Prior composition}
As described in Sect. \ref{sec:StructureOfAlgorithm}, we consider two different prior models, one for the \ac{SF} reconstruction and one for the \ac{MF} reconstruction. The fields in the \ac{SF} reconstruction are defined in the spatial domain, $s_\text{SF}:\Omega_{\text{SF}} = \mathbb{R}^2\to\mathbb{R}^+$, whereas the \ac{MF} fields have an additional spectral dimension, $s_\text{MF}:\Omega_{\text{MF}}=\mathbb{R}^2 \times \mathbb{R}\to\mathbb{R}^+$.  Regardless of the model, we assume that the X-ray sky consists of two possible sources: point sources and diffuse sources. Different prior models represent fluxes of different morphologies, each shaped by their physical production processes. The flux in diffuse structures should vary smoothly over position space. In other words, field values in the vicinity of a location are similar to that, which is best represented by the correlation structure of the field. In contrast, point sources are spatially uncorrelated and therefore best represented by spatially independent and sparsity enforcing priors. We discuss either component in more detail below. In the following, the validity of the assumptions made for $s\in \{s_\text{SF}, s_\text{MF}\}$ is assumed to hold for both the \ac{SF} and \ac{MF} sky.\\
We represent the flux signal $s$ as a superposition of point sources $s_\text{p}$ and diffuse sources $s_\text{d}$. In addition, we add a background component $s_\text{b}$, which in our case accounts for the different backgrounds in \ac{FI} and \ac{BI} chips, which are further discussed in Sect. \ref{sec:InstrumentResponse}. \\
 Correspondingly, we denote the latent space sub-vectors, which parametrize these individual components, as $\xi_\text{p}$, $\xi_\text{d}$, and $\xi_\text{b}$, which in composition form the total latent space vector of the model $\xi = (\xi_\text{p}, \xi_\text{d}, \xi_\text{b})$. The according generative prior model is given by,
\begin{align}
\label{eq:signal_superposition}
 s(\xi) = \underbrace{s_\text{p}(\xi_\text{p}) + s_\text{d}(\xi_\text{d})}_{\text{sky flux}} + \underbrace{ R^\prime s_\text{b}(\xi_\text{b})}_\text{\ac{BI} background}.
\end{align}
Here, $R^\prime$ denotes a mask, which assures that the additional background field is added in \ac{BI} chip regions only. 
By expressing the transformation into the standardized coordinate system as a function $s_i$ with $i \in {\text{d, p, b}}$, we obtain a generative model for each component as described in \citet{En_lin_2022}. \\
A set of prior samples can be seen in the first row of the synthetic data generation diagram in Fig. \ref{fig:priors} in the appendix. Furthermore, the according prior samples and synthetic data allow us to choose the model hyper-parameters correctly. Here, we perform the search in two steps. First, we ensure mathematically via a coarse adjustment of the parameters that the order of magnitude in the counts of the data in Fig. \ref{fig:data} is the same as the order of magnitude in the expected counts of the pixel-wise product of the exposure and the prior samples. Second, we look at the corresponding synthetic data, as shown in Fig. \ref{fig:priors} and fine-tune hyper-parameters such that the components in the actual data and the synthetic data are morphologically
similar, in order to improve the convergence of the algorithm. 
\subsection{Correlated components}
Correlated components correspond to a flux that can vary over several orders of magnitude and exhibit spatial correlation. In this sense, the diffuse sky emission and the background are represented by correlated components.  Their morphology is implemented by representing the signal for a correlated component as a log-normal processes,
\begin{align}
 s = e^{\tau} ~~\text{with}~~\mathcal{P}(\tau|T) = \mathcal{G}(\tau, T),
\end{align}
with an unknown covariance $T$ describing the correlation structure of the correlated signal component.  Since the correlation structure is not known a priori, we infer it concurrently by incorporating the correlated field model from \citet{Arras_2022}. Using the reparametrization trick introduced by \citet{kingma2015variational}, we describe the logarithmic sky flux as a generative process,
\begin{align}
\label{eq:gen_process_corr_field}
\tau= A\xi_{\tau} ~~ \text{with} ~~T = A A^\dagger.
\end{align}
We model the correlations in space and energy separately and assume a priori statistical homogeneity and isotropy of the correlated logarithmic sky flux components in each of the subspaces $\Omega^{(k)}$, where $\Omega=\bigotimes_k \Omega^{(k)}$. Thus, according to the Wiener-Khinchin theorem, the corresponding covariances for the space $\Omega^{(k)}$, $T^{(k)}$, are diagonal and defined by the power spectrum $p_{T^{(k)}}$. \\
To learn the correlation structure of the correlated component, the power spectrum is modeled non-parametrically by representing the logarithmic power spectrum by an integrated Wiener process according to \citet{Arras_2022}. In particular, the mean and uncertainty of the parameters resulting from the chosen representation, such as the slope of the logarithmic power spectrum, its offset and the fluctuations around the described power law, are learned from the data by modeling them as generative processes. This introduces further latent parameters to describe the generative model for the correlation structure. In the following, we will refer to this prior model as the correlated field. \\
In case of the \ac{SF} model we only consider spatial correlations. Accordingly, the generative model for the spatially correlated components in the SF reconstruction is defined via $s_{\text{SF}}(\xi_\text{SF}) = e^{\tau_\text{SF}(\xi_\text{SF})}$ with $\Omega=\Omega^{(k)}=\mathbb{R}^2$.
In the \ac{MF} model we combine the power spectra for the independent spatial and spectral domain via a tensor product and define $s_{\text{MF}}(\xi_\text{MF}) = e^{\tau_\text{MF}(\xi_\text{MF})} $. A further description of this generative model and its normalization can be found in \citet{Arras_2022}. \\
The diffuse and background components are represented by these spatially and spectrally correlated components. The number of derived hyper-parameters as well as latent parameters per component in each model is given in Table \ref{tab:parameter_table} and Table \ref{tab:latent_parameter_table}. For the correlation structure of the diffuse and background components, we make different prior assumptions to ensure an adequate separation of these components. In particular, we assume that the spatial power spectrum of the diffuse sky structures has a slightly declining slope, allowing for small-scale structures in this component, while the spectrum of the background is assumed to be steep, allowing only for smooth background noise in the back-illuminated chips.
\subsection{Point-like components} 
Point-like components appear local without any spatial correlation structure, due to their extreme distances. Consequently, we  assume that the sky fluxes from point sources are spatially independent and thus their prior factorizes in spatial direction,
\begin{align*}
\mathcal{P}(s_{p})=\prod_{\vec{x}} \mathcal{P}(s_p(\vec{x},y)).
\end{align*}
In \citet{Selig_2015} different functional forms of possible point source luminosity priors were analyzed. Since the reconstruction of SN1006 requires a point source prior capable of modeling a few very bright point sources, we choose the inverse-gamma prior for the spatial direction according to \citet{Guglielmetti_2009},
 \begin{align}
 \label{eq:InverseGamma}
 \mathcal{P}(s_p|q,\alpha) = \prod_x \frac{(q)^{\alpha}}{\Gamma(\alpha)}\biggl(\frac{1}{s_p^{(x)}}\biggr)^{\alpha+1}\exp\biggl(\frac{-q}{s_p^{(x)}}\biggr),
 \end{align}
where $\alpha$ is the shape parameter of the inverse-gamma distribution and $q$ is the corresponding lower flux cutoff. The inverse-gamma prior behaves as a power law for fluxes much larger than the cutoff value,
which matches the behavior observed for the luminosity functions of high-energy astrophysical point sources.
Intuitively, it encodes the assumption that with increasing luminosity, the set of point sources exceeding it becomes increasingly sparse. We model the inverse-gamma prior in standardized coordinates via inverse transform sampling leading to the generative model,
\begin{align}
\label{eq:SF_points}
s_\text{SF}=s_p(\xi_p),
\end{align}
where $\xi_p$ is drawn from a standard Gaussian. Accordingly, $s_p$ encodes the entire complexity of the inverse-gamma distribution and $s_p(\xi_p)$ is drawn according to Eq.~\eqref{eq:InverseGamma}.
In the \ac{SF} model, Eq.~\eqref{eq:SF_points} describes the accurate generative model for the point sources. \\
For the \ac{MF} model, we need to consider the spectral axis as well, by modeling the point source flux as spatially independent functions of the logarithmic energy $y$, according to \citet{platz2022multicomponent}. We assume that each point in the spatial subdomain has non-negligible correlations in the energy direction, as described by the correlated field component. In particular, we want to obtain a power law dependence in the energy direction, defined by the spectral index $a$, and add fluctuations around it by a correlated field $\tau_p$,
 \begin{align}
 \label{es:MF_points}
 (s_\text{MF})_p(x,y) = (s_\text{SF})_p(x)  \frac{e^{\tau_p^{(x)}(y)+a_p^{(x)} y}}{C},
 \end{align}
where $C$ is the normalization. Here, not only the correlated field is described by a generative model but also the spectral index $a_p^{(x)}$ at every location $x$ is learned. Thus, the additional energy axis introduces a number of new hyper- and latent parameters. The exact numbers of hyper- and latent parameters for the point sources in each model are given in Table \ref{tab:parameter_table} and Table \ref{tab:latent_parameter_table}.
\begin{table}
\centering
\begin{tabular}{ccccc}\toprule
\bfseries Model & \bfseries $s$ & \bfseries $s_d$ & \bfseries $s_p$ & \bfseries $s_b$ \\\midrule
\bfseries \ac{SF} &  24 & 11 & 2 & 11 \\
\bfseries \ac{MF} & 53 & 19 & 15 & 19 \\\bottomrule
\end{tabular}
\caption{Number of hyper-parameters in each model per component}
\label{tab:parameter_table}
\end{table}

\begin{table}
\centering
\begin{tabular}{ccccc}\toprule
\bfseries Model & \bfseries $s$ & \bfseries $s_d$ & \bfseries $s_p$ & \bfseries $s_b$ \\\midrule
\bfseries \ac{SF} &  $3.4 \times 10^6$ & $1.2 \times 10^6$ &  $1.0 \times 10^6$  &  $1.2 \times 10^6$  \\
\bfseries \ac{MF} & $1.4 \times 10^7$ & $4.4 \times 10^6$   & $5.0 \times 10^6$  & $4.4 \times 10^6$  \\\bottomrule
\end{tabular}
\caption{Number of latent parameters in each model per component}
\label{tab:latent_parameter_table}
\end{table}

\section{Chandra instrument and data description}
\label{sec:InstrumentResponse}
In Bayesian X-ray imaging, the prior model (Sect. \ref{sec:Priors}) is responsible for decomposing the components, whereas the denoising and deconvolution is controlled by the likelihood model, which describes the measurement process. In general, an X-ray telescope provides photon counts that are statistically binned into pixels. This stochasticity is modeled by Poisson noise. The Poisson distribution gives the probability of the actual number of photon counts per bin, given the expected number of events, $\lambda$,
\begin{align}
\label{eq:Likelihood}
\mathcal{P}(\vec{d}|\vec{\lambda})=\prod_i \mathcal{P}(d_i|\lambda_i) = \prod_i \frac{1}{d_i!}{\lambda_i}^{d_i}e^{-\lambda_i}.
\end{align}
In the end, we want to know the photon flux at each point in position and energy space. To do this, we need to model the response function $R$, which in a first step transforms the continuous flux field into a pixel-wise vector of expected photon counts $\lambda_i$, given a sky and \ac{BI} background model. The response function includes all aspects of the instrument specific measurement, which are described in more detail below. Given the response $R$ (Sect. \ref{subsec:response}), the number of expected counts at each pixel, $\lambda_i$, is calculated via $\vec{\lambda}(z) = R(s)(z)$.
Because the Chandra \ac{FOV} is small compared to the  extent of SN1006, multiple observations were taken to cover the whole \ac{SNR}. For each of several data patches $j$, we get the data $\vec{d}_j$ and the response $R_j$, which need to be fused. Here we introduce a mechanism that accounts for differences in the exposure and the \ac{PSF} between the patches. By assuming that each patch is observed independently, we can write the log-likelihood of the full mosaic as the sum over individual patches (Eq.~\eqref{eq:Likelihood}) for each data patch $\vec{d}_j$ and the corresponding expected counts $\vec{\lambda}_j$ calculated from the response $R_j$,
\begin{align}
\label{eq:Loglhsum}
\ln \mathcal{P}(\vec{d}|\vec{\lambda}) = \sum_j \ln \mathcal{P}(\vec{d}_j|\vec{\lambda}_j).
\end{align}
\subsection{Chandra instrument response}
\label{subsec:response}
We consider the data taken by the \ac{ACIS} \citep{2003SPIE.4851...28G}, which is able to determine the energy of each incoming photon by using \ac{CCDs}. In particular, we consider the energy range 0.5 keV to 7.0 keV, which we bin in accordance with \citet{winkler2014high} into three energy bins (0.5-1.2 keV, 1.2-2.0 keV, 2.0-7.0 keV). Chandra carries two different kinds of \ac{ACIS} detectors, ACIS-I which is used for imaging and ACIS-S which is used for imaging and spectral analysis. According to their usage ACIS-I and ACIS-S differ in the chips they are built of. In particular, ACIS-I is constructed out of \ac{FI} chips only, which means that the incidental X-ray photons have to pass through the metal wiring until they reach the light receiving surface. In contrast ACIS-S also contains \ac{BI} chips, where the CCD is flipped, such that the gate structure and channel stops do not face the X-ray illuminated side. Accordingly, the \ac{BI} chips are more sensitive to soft X-rays and thus are well suited for spectral analysis purpose. However, they have a lower high-energy quantum efficiency and a worse resolution due to increased noise  \citep{arnaud_smith_siemiginowska_2011}. The exact layout of ACIS-I and ACIS-S can be found in \citet{ChandraPOG}.\\
We use version 4.14 of \ac{CIAO} tool \citep{CIAO} designed by the \ac{CXC} to extract information on the response ingredients like the \ac{PSF} and the exposure as well as on the event files itself for each patch. Here, we make use of tools from the the category of "Data Manipulation" for extracting and binning the data and from the category "Response Tool" to generate the ingredients of the instrument response. 

\begin{figure}
\centering
\begin{subfigure}{\linewidth}
  \centering
  \includegraphics[width=\linewidth]{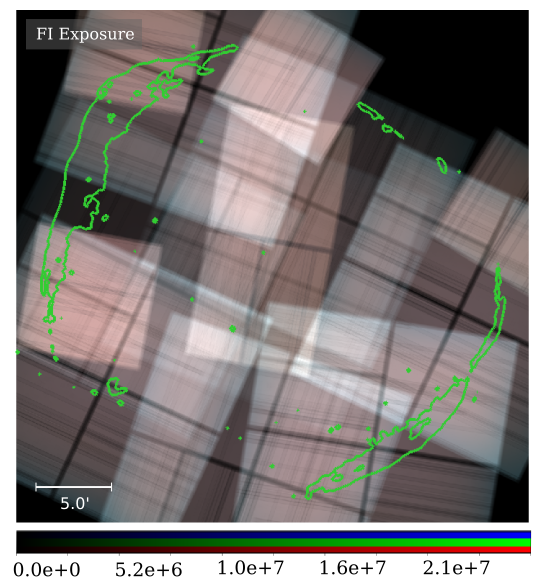}
\end{subfigure} \\
\begin{subfigure}{\linewidth}
  \centering
  \includegraphics[width=\linewidth]{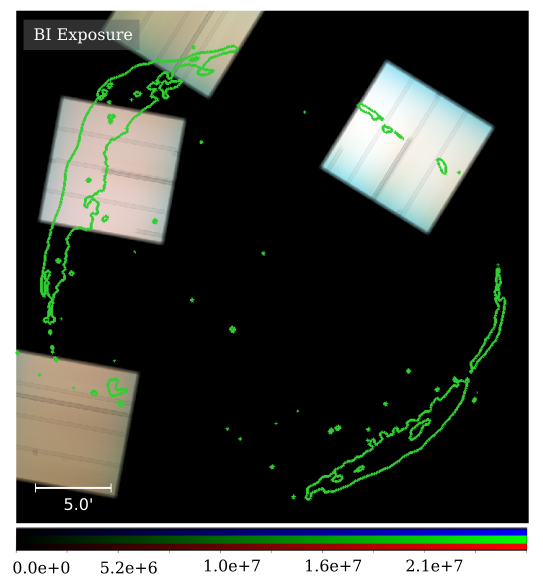}
\end{subfigure} 
\caption{Visualization of the exposures used for the reconstruction in $[\text{s cm}^2]$(Table \ref{tab:data_table}): Top: Full exposure for all \ac{FI} chips. Bottom: Full exposure for all \ac{BI} chips.}
\label{fig:exposure}
\end{figure}

The exposure map is a key component in the process of converting raw X-ray data into scientifically useful data products, such as images and spectra. The exposure map combines information from the instrument map, which characterizes the instrument sensitivity like the effective area, and the aspect solution \citep{X-RayAnalysis}, which describes the spacecraft pointing and roll, to create a map of the total observing time, or exposure, for each pixel in the field of view.\\
In \citet{effective_area} the effective area as a function of energy is shown for the different chips. As mentioned above already, the \ac{FI} chips are much less sensitive to low energy X-ray photons than the \ac{BI} chips. On the other hand, the \ac{BI} chips have more background flux. The exposure maps for the \ac{FI} and \ac{BI} chips can be seen in Fig. \ref{fig:exposure}. In order to account for the higher noise in the \ac{BI} chips, we introduced an additional \ac{BI} background field in Sect. \ref{sec:Priors}.\\
In general, for any of the considered chip types it can be seen that the chips are more sensitive for higher energies, leading us to the decision to take the highest energy bin (2.0-7.0 keV) as a starting point for the transition model.
The \ac{PSF} was simulated using the \ac{MARX} \citep{MARX}. \ac{MARX} is a software developed by the \ac{CXC} amongst others to simulate the response, more exactly the \ac{PSF}, of the Chandra X-ray Observatory taking into account the telescope optics, the pointing and the aspect of the telescope. We generate the response for each dataset and thus use for each dataset a homogeneous, spatially invariant \ac{PSF} but different \ac{PSF}s for the different patches. The consequences of this approximation are addressed further in the quantitative discussion of the results in Sect. \ref{sec:QuantitativeDiscussion}. For a further analysis of spatially variant \ac{PSF}s we refer to \citet{e25040652}. 
\subsection{Chandra data of SN1006}
The photon count data taken by the instrument is translated into an event table. Each event has information on time, energy and position. Here, the data is binned into $1024 \times 1024$ spatial pixels and three energy bins. Moreover, the pointing direction of Chandra varies in time.  Thus  aspect correction, meaning taking into account the pointing direction of the telescope as a function of time is necessary. The data itself is an event list, which specifies the position in the chip coordinates and the arrival time of each photon. In \citet{mcdowell2001coordinate} the calculation of the sky coordinates of the photon given this event list is specified.\\
The latest Chandra observations of SN1006 according to \citet{winkler2014high} was chosen for the here presented reconstruction. The information on the data is listed in Table \ref{tab:data_table}. The aim  of the study of \citet{winkler2014high} was to measure the motion  of the remnant and to get a more detailed view on its fine scale structures. Thus, the data observations were designed by \citet{winkler2014high} in order to match former observations, to be able to measure the expansion and using a longer exposure time in order to get a more detailed picture. The previous observations, which were taken as first-epoch images by \citet{winkler2014high} studied the non-thermal NE rim and the thermal NW rim with the ACIS-S array \citep{Long_2003, Katsuda_2009} and the whole remnant with a mosaic of ACIS-I observation \citep{Cassam_Chena__2008}. Accordingly, the reconstruction is dealing with the data from \ac{BI} and \ac{FI} chips.

\begin{figure}
  \centering
  \includegraphics[width=\linewidth]{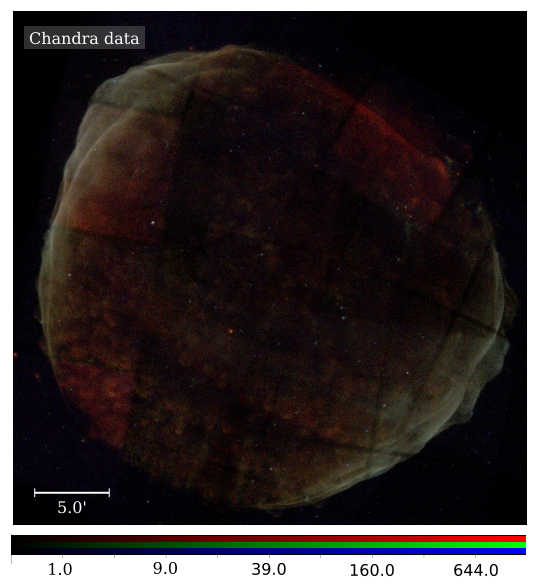}
  \caption{Visualization of the photon count data used for the reconstruction (Table \ref{tab:data_table}) with right ascension on the x-axis and declination on the y-axis: red = 0.5-1.2 keV, green = 1.2-2.0 keV, blue = 2.0-7.0 keV.}
  \label{fig:data}
\end{figure}

\begin{table}
\centering
\begin{tabular}{ccccc} \toprule
\bfseries ObsID & \bfseries Instrument & \bfseries R.A. & \bfseries Decl. & \bfseries Date\\ \midrule
9107 & ACIS-S & 15:03:51.5 & -41:51:19 & 24.06.2008 \\
13737 & ACIS-S & 15:02:15.9 & -41:46:10 & 20.04.2012 \\
13738 & ACIS-I & 15:01:41.8 & -41:58:15 & 23.04.2012\\
14424 & ACIS-I & 15:01:41.8 & -41:58:15 & 27.04.2012\\
13739 & ACIS-I & 15:02:12.6 & -42:07:01 & 04.05.2012 \\
13740 & ACIS-I & 15:02:40.7 & -41:50:21 & 10.06.2012 \\
13741 & ACIS-I & 15:03:48.0 & -42:02:53 & 25.04.2012 \\
13742 & ACIS-I & 15:03:01.8 & -42:08:27 & 15.06.2012 \\
13743 & ACIS-I & 15:03:01.8 & -41:43:05 & 28.04.2012 \\
14423 & ACIS-I & 15:02:50.9 & -41:55:25 & 25.04.2012 \\
14435 & ACIS-I & 15:03:42.5 & -41:54:49 & 08.06.2012 \\\bottomrule
\end{tabular}
\caption{Information on the Chandra ACIS observations for the used data of SN1006 according to \citep{winkler2014high}. Observations taken by the instrument ACIS-S are followed by the ACIS-I observations. 
}
\label{tab:data_table}
\end{table}
\section{Validation of the algorithm using synthetic data}
\label{sec:mock}
\begin{figure*}[ht!]
  \centering  
 \includegraphics[width=\textwidth]{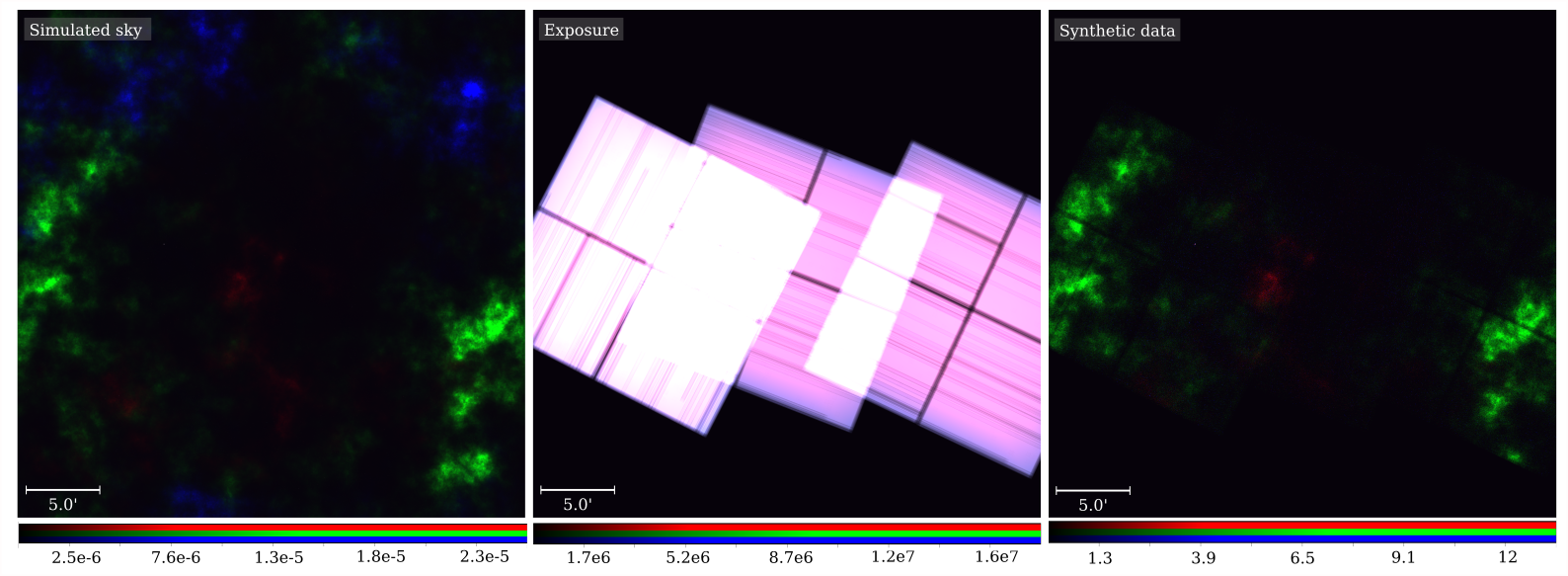}
  \caption{Generation of synthetic data: Left: Sky sample generated for the validation experiment. Center: Chandra exposure, modeled by combining three patches (14423, 14424, 14435) (see Table \ref{tab:data_table}). Right: Synthetic data corresponding to the sky sample, obtained by convolving the sky sample with the \ac{PSF} and drawing a pixel-wise Poisson sample from the resulting detector flux prediction.}
  \label{fig:mock_data_generation}
\end{figure*}

\begin{figure*}[ht]
  \centering  
  \centering  
 \includegraphics[width=\textwidth]{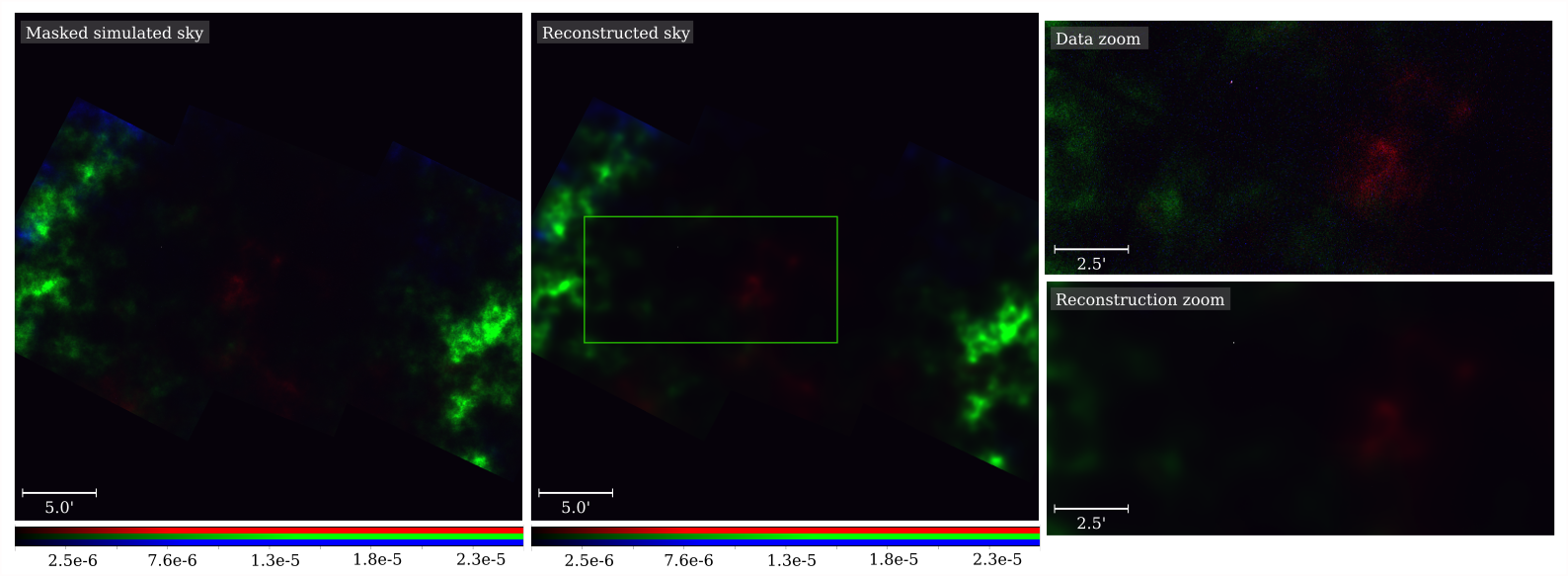}
  \caption{Reconstruction results on synthetic data: Left: Sky sample generated for this study masked by the extent of the exposure patches (14423, 14424, 14435) (see Table \ref{tab:data_table}). Center: Reconstruction result, i.e. the posterior mean, of the imaged sky masked by the extent of the same exposure patches. Right: Zoomed regions of the data on top and of the reconstructed image below. The shown cutout region is marked in the center image.}
    \label{fig:mock_reconstruction}
\end{figure*}

\begin{figure}
\centering
    \includegraphics[width=1.0\linewidth]{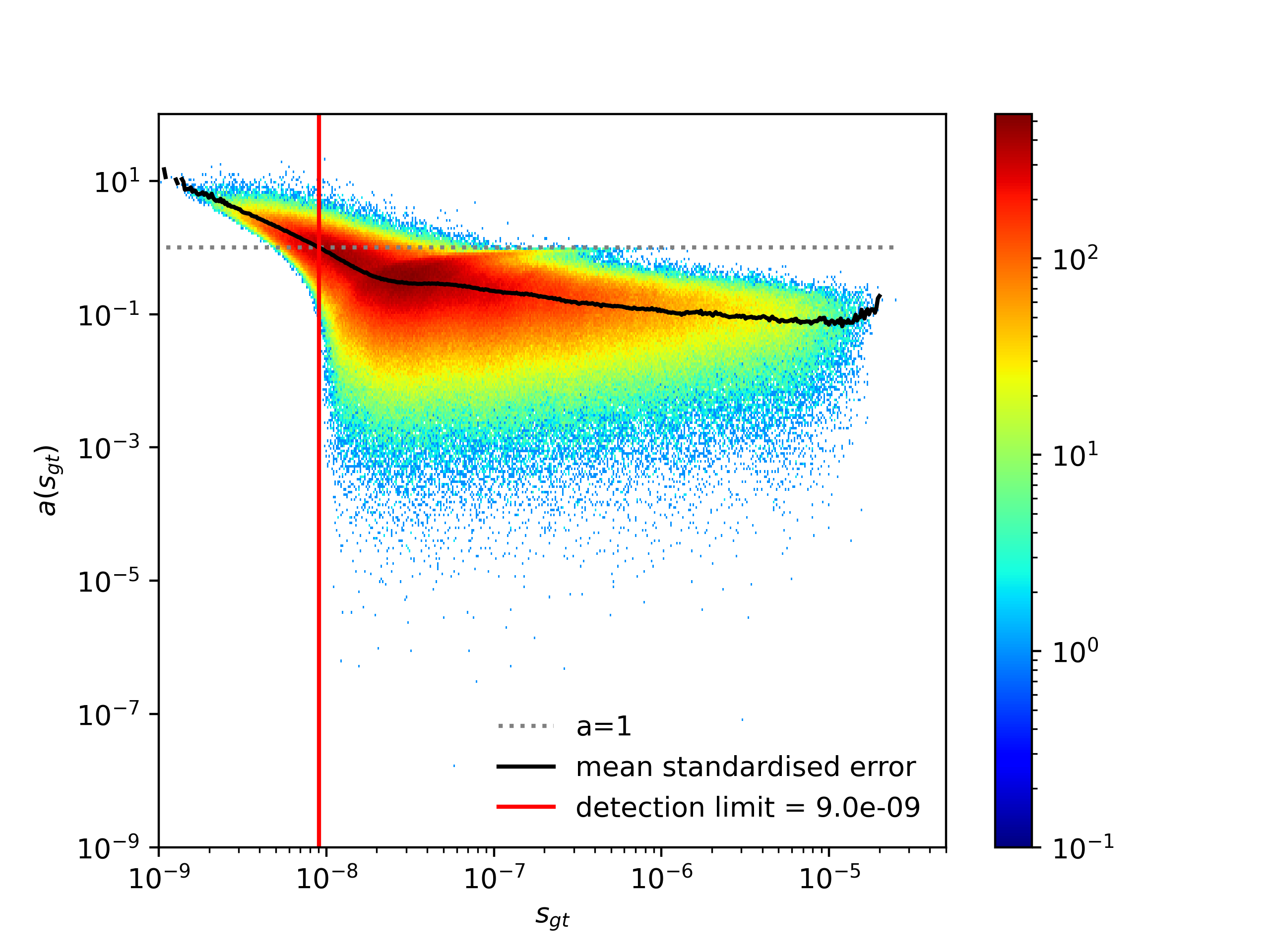}
    \caption{Visualization of the two-dimensional histogram for the sample averaged relative distance of the posterior sky flux samples vs. the ground truth sky flux (Eq.~ \eqref{eq:relative_sample_distance}). The  detection limit is determined via the intersection of the line showing the mean standardized error $\bar{a}$ (Eq.~\eqref{eq:integrated_counts}) with the a=1 line.}
    \label{fig:relfluxdiagnostics}
\end{figure}

To demonstrate the performance of the developed algorithm, we perform the inference described above on a realistic but simulated dataset. Such a reconstruction on synthetic data is useful not only for validating the reconstruction method, but also for determining certain parameters of the actual reconstruction, such as the sky flux detection limit. By constructing our model as a generative model, we are able to draw realistic sky samples that are similar to the region of the X-ray sky considered in this study. Given a sample of the sky, we can apply the response to it and mimick Poisson noise. As a result we get synthetic data. The process of generating synthetic data is illustrated in the Appendix \ref{sec:synthdata}.
For simplicity, we consider only three of the ACIS-I exposure patches for the synthetic reconstruction, rather than the whole set. The according synthetic data is shown in Fig. \ref{fig:mock_data_generation}, together with the actual simulated sky sample and the considered exposure map.  \\
The resulting spatio-spectral reconstruction of the synthetic data in Fig. \ref{fig:mock_reconstruction} shows that the structures of the simulated sky are well captured. The denoising and response corrections are clearly visible when compared to the data shown in Fig. \ref{fig:mock_data_generation}. In particular, the right-hand side Fig. \ref{fig:mock_reconstruction} shows an enlarged version of the data and reconstruction to illustrate the denoising and deconvolution. \\
As mentioned before, a particular strength of the X-ray imaging method presented here is that we not only get an expectation of the signal, but also a corresponding standard deviation to this estimate. The corresponding pictures of the standard deviation for the individual energy bins can be seen in the appendix in Fig. \ref{fig:mockskyuncertainty}. As expected, higher mean values exhibit greater variability in the flux, which in turn leads to a higher absolute uncertainty. However, given the standard deviation $\sigma_s$, the reconstruction mean $m_s$ and the fact that we know the signal ground truth $s_{\text{gt}}$ itself from our generative model, we can calculate even more interesting validation measures, like the \ac{UWRs} per energy bin $i$,
\begin{align}
(\epsilon_\text{UWR})_i = \frac{(m_s)_i-(s_\text{gt})_i}{\sigma_i}.
\label{eq:uncertaintyweightedresiudal}
\end{align}
Fig. \ref{fig:mockskyuwr} in the appendix shows the \ac{UWRs} as well as the residuals,
\begin{align}
r = (m_s - s_\text{gt}).
\label{eq:Residuals}
\end{align}
Areas with many counts, show higher and more correlated residuals than areas with low counts. Accordingly, the \ac{UWRs} show that these pixels with high counts have higher uncertainty weighted residuals, due to a relatively small uncertainty. Overall, the simulated reconstruction demonstrates that the method developed is internally consistent. Therefore, we use this synthetic reconstruction to set the threshold above which we can no longer distinguish noise from signal, which we will refer to as the detection limit. The detection limit is used as a plotting lower limit in the actual reconstruction. Below this lower limit flux values are not shown in the image.\\
The posterior approximation gives us the opportunity to draw posterior samples $s^* \hookleftarrow \mathcal{Q}(s|d)$, with which we can calculate sample averages. In order to determine the detection limit, we define the standardized error, $a$, between the ground truth $s_\text{gt}$ and each of these samples $s^*$ as a function of the ground truth flux,
\begin{align}
\label{eq:relative_sample_distance}
a(s_\text{gt}) = \bigg \vert \frac{s^*-s_\text{gt}}{s_\text{gt}}\bigg \vert.
\end{align}
In Fig. \ref{fig:relfluxdiagnostics}, the sample-averaged two-dimensional histogram $a(s_\text{gt})$ as a function of the ground truth flux $s_\text{gt}$ is shown. For each value $s_\text{gt, i}$ we calculate the mean standardized error of the histogram bins along the $a$-axis, $\bar{a}(s_\text{gt, i})$, where $n(a_j, s_\text{gt, i})$ is the number of counts for each bin $(i, j)$ in the two-dimensional histogram,
\begin{align}
\label{eq:integrated_counts}
\bar{a}(s_\text{gt, i}) = \frac{\sum_j a_j *n(a_j, s_\text{gt, i})}{\sum_j n(a_j, s_\text{gt, i})}.
\end{align} 
Fig. \ref{fig:relfluxdiagnostics} reflects the expectation that the standardized error increases with smaller flux. To establish a detection threshold for low fluxes, we define a limit beyond which we cannot confidently ascertain the presence of a signal in our observations. In this study, we determine the detection threshold such, that the mean standardized error is less than one. In other words, the detection threshold is set via the intersection point of the mean standardized error $\bar{a}$ and the line $a=1$, leading to the detection limit $9.0e^{-9} [\text{s}^{-1} ~\text{cm}^{-2}]$. We only performed this synthetic analysis on three of the data patches. Therefore, this threshold is conservative for the reconstruction with all patches.

\section{Results and analysis of inference performance}
\label{sec:results}
\begin{figure}[ht]
  \centering  
    \includegraphics[width=\linewidth]{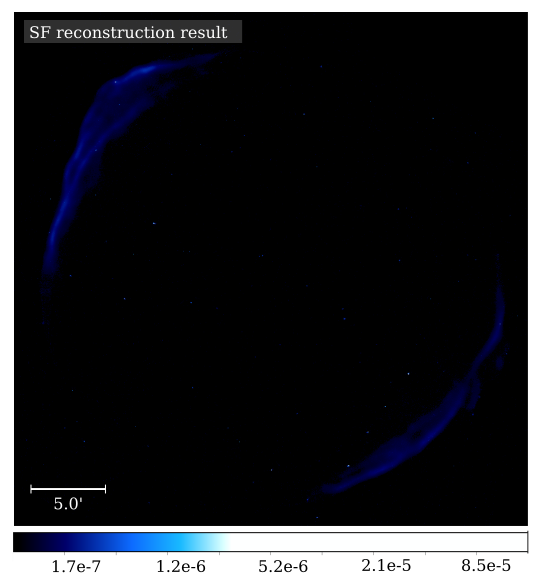}
    \caption{Spatial reconstruction result for the highest energy bin in $[\text{s}^{-1}~\text{cm}^{-2}]$.}
    \label{fig:singelf_reconstruction}
\end{figure}

\begin{figure*}
  \centering
  \begin{subfigure}[t]{\textwidth}
    \centering
    \includegraphics[width=0.8\textwidth]{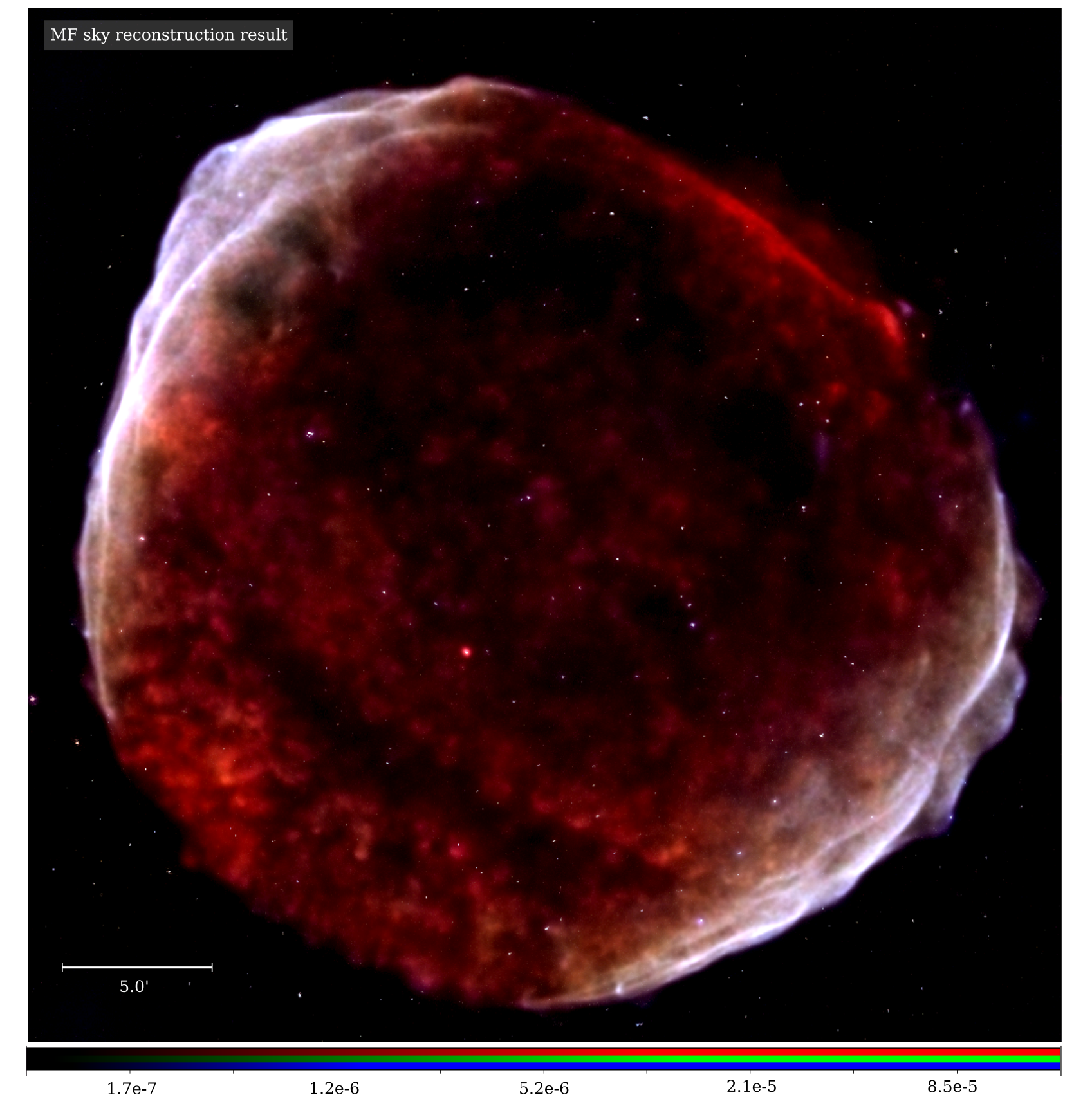}
  \end{subfigure} \\
  \hspace{-2.5cm}
  \vspace{-0.5cm}\\
    \begin{subfigure}[t]{0.45\textwidth}
    \includegraphics[width=\textwidth]{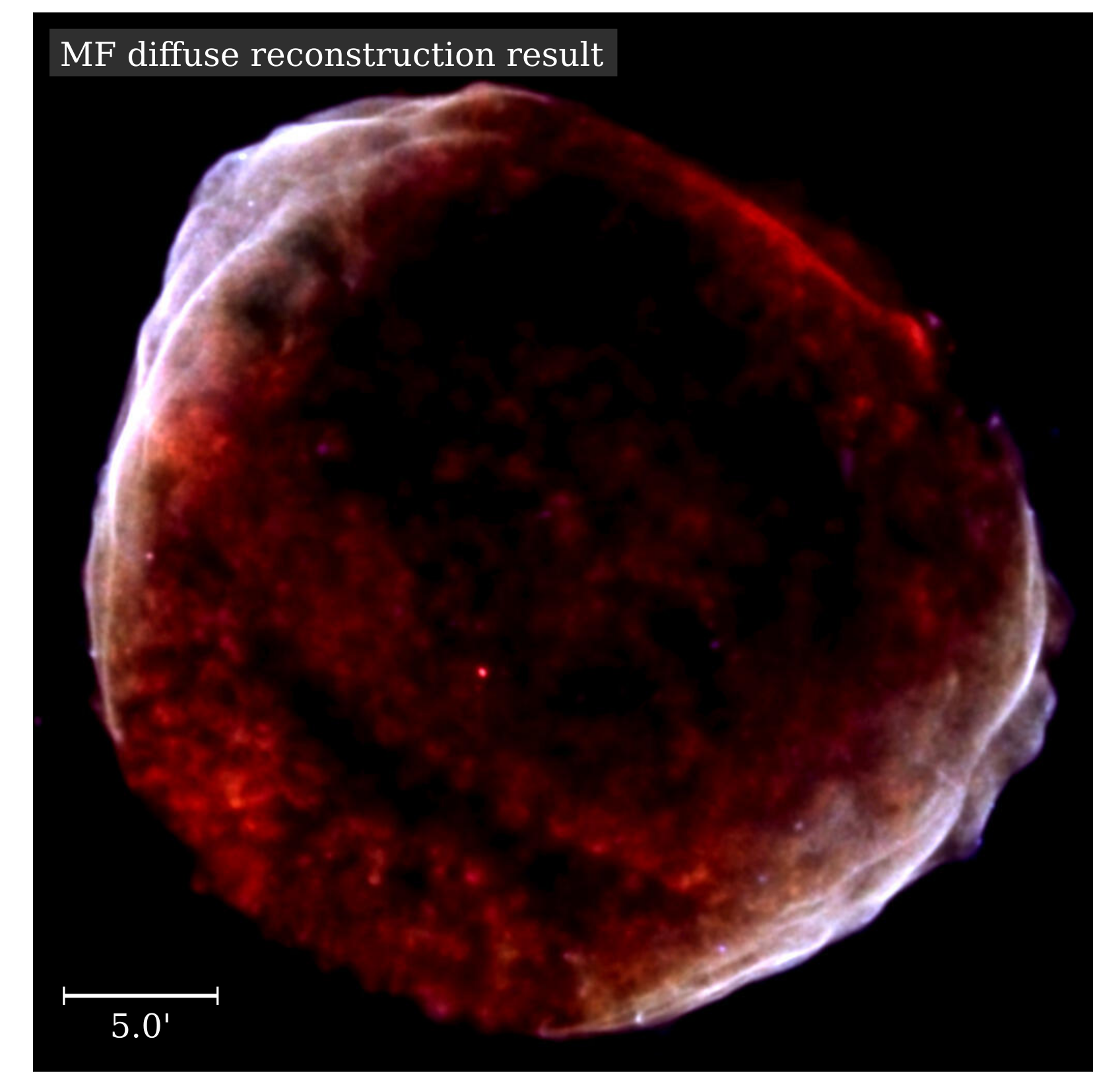}
  \end{subfigure}
  \begin{subfigure}[t]{0.45\textwidth}
    \includegraphics[width=\textwidth]{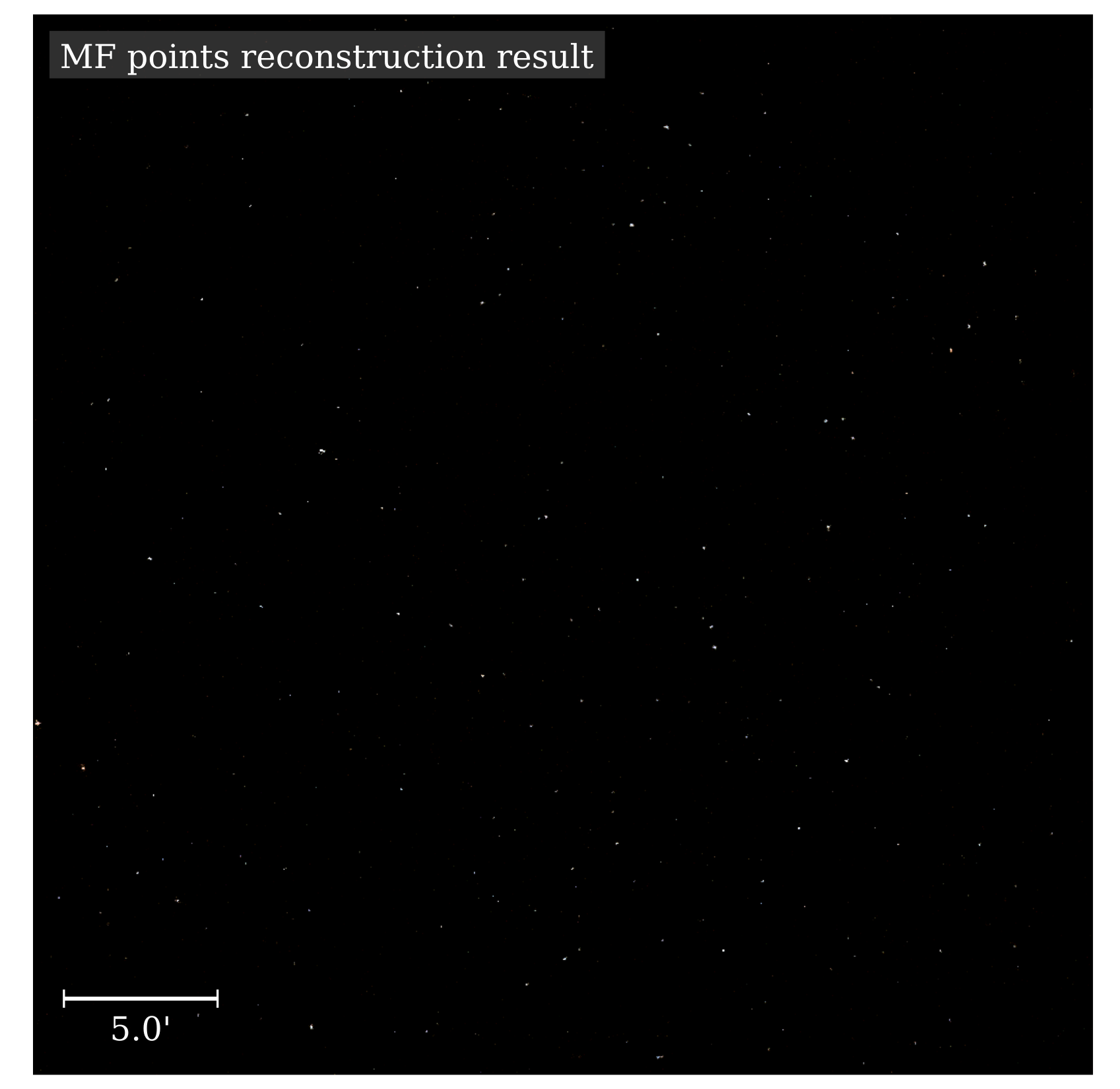}
  \end{subfigure}
  \caption{Reconstruction results for the flux in $[\text{s}^{-1}~\text{cm}^{-2}]$ (red = 0.5-1.2 keV, green = 1.2-2.0 keV, blue = 2.0-7.0 keV. The corresponding color bars for each of these energy bins can be found in Fig. \ref{fig:uncertainty}): Top: Full sky reconstruction mean. Bottom left: Reconstruction mean for diffuse emission. Bottom right: Reconstruction mean for point sources.}
  \label{fig:reconstruction_result}
\end{figure*}

In this section we discuss the results of the sky flux reconstruction. The additional background from the \ac{BI} chips according to Eq.~\eqref{eq:signal_superposition} is removed in the reconstruction process.
In Fig. \ref{fig:singelf_reconstruction} we show the intermediate result of the \ac{SF} reconstruction of the highest energy bin. This is taken as the initial condition of the subsequent \ac{MF} reconstruction of SN1006, whose reconstruction results are shown in Fig. \ref{fig:reconstruction_result} given the data shown in Table \ref{tab:data_table} using Bayesian imaging and the transition model introduced in Sect. \ref{sec:StructureOfAlgorithm}. The reconstruction was visualized using the SAOImage DS9 imaging application
\citep{2003ASPC..295..489J}. The according results for each energy bin including the according color bars can be found in the appendix in Fig.\ref{fig:uncertainty}. \\
As mentioned, we are not only reconstructing the sky flux itself but also its correlation structure in its correlated components. Accordingly, the posterior mean of the power spectrum in the spatial direction of the diffuse, extended sky component was also reconstructed and is shown for each energy bin in the appendix in Fig. \ref{fig:spatial_pspec}.

\begin{figure}
  \centering  
    \includegraphics[width=\linewidth]{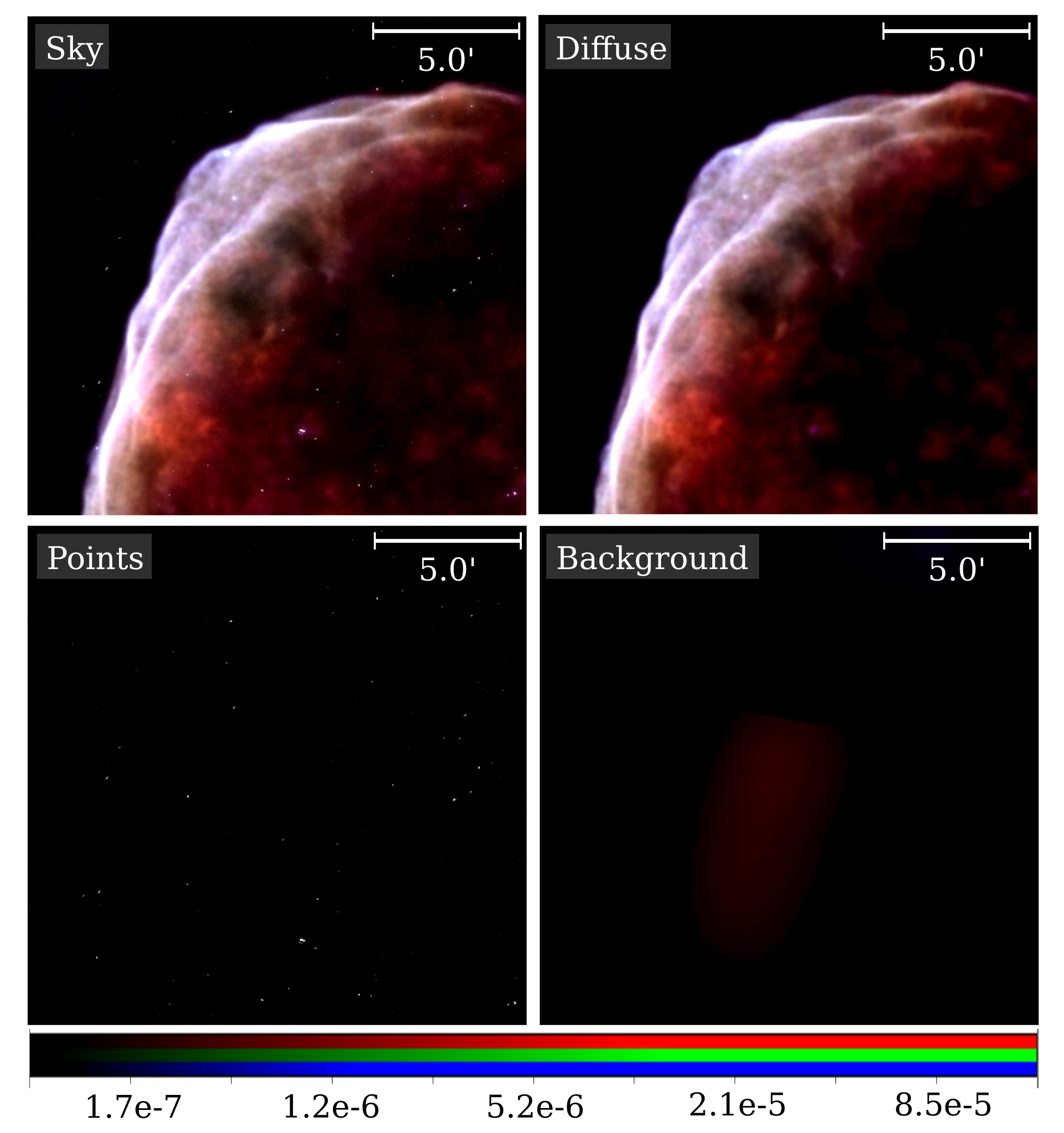}
    \caption{Northeastern quarter of SN1006 and its components: From left to right: total sky, diffuse emission, point sources, \ac{BI} background.}
    \label{fig:Remnantzoom}
\end{figure}
\begin{figure}
  \centering  
    \includegraphics[width=\linewidth]{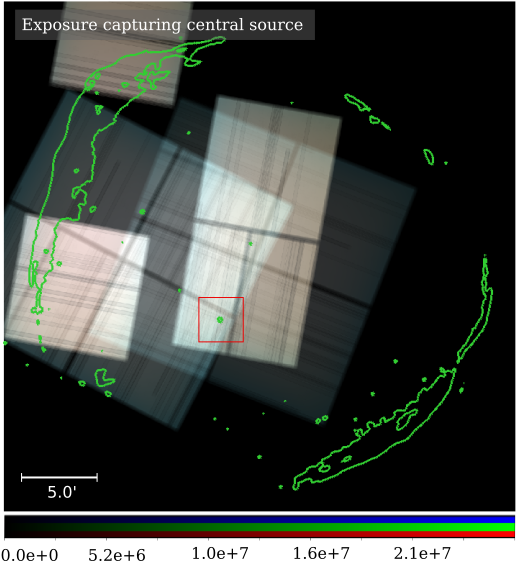}
    \caption{Exposures that capture the not well separated point source (marked red).}
    \label{fig:psf_exposure}
\end{figure}
\begin{figure}
\centering
\begin{subfigure}[b]{\linewidth}
\centering
    \includegraphics[width=\linewidth]{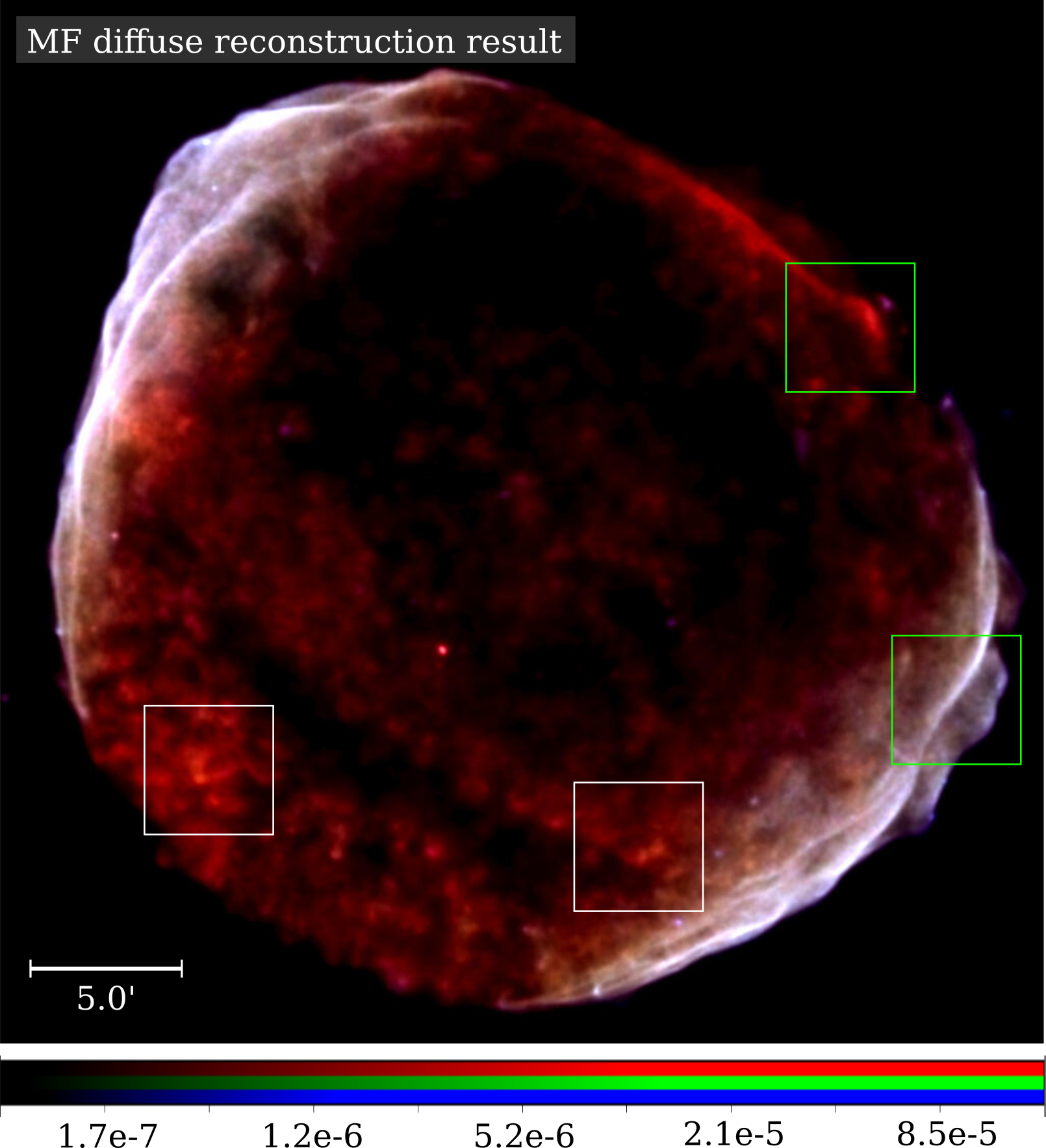}
    \caption{}
  \end{subfigure} \\
\begin{subfigure}[b]{\linewidth}
\centering
    \includegraphics[width=\linewidth]{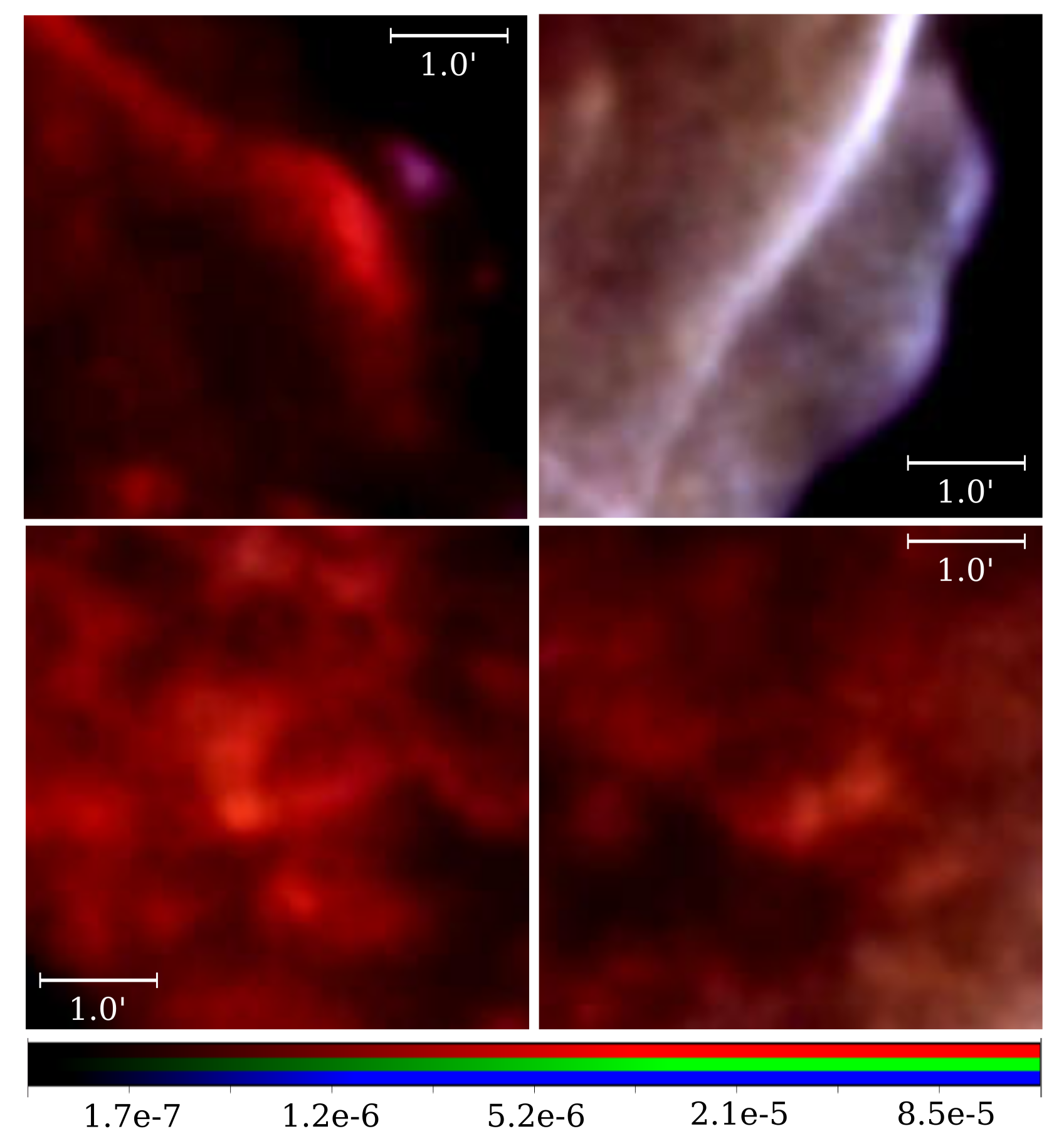}
    \caption{}
  \end{subfigure} \\
  \caption{Zoom in of the reconstruction results in the diffuse component: (a) The whole diffuse reconstruction and the locations of the zoomed areas. (b) The top panels represent the green areas marked in the remnant and show zooms on the denoised shell of the remnant. The lower panels are represented by the white boxes in (a) and show structures in the inner soft X-ray emission of the remnant.}
  \label{fig:smallzooms}
\end{figure}
\begin{figure*}
  \centering
  \includegraphics[width=0.3\textwidth]{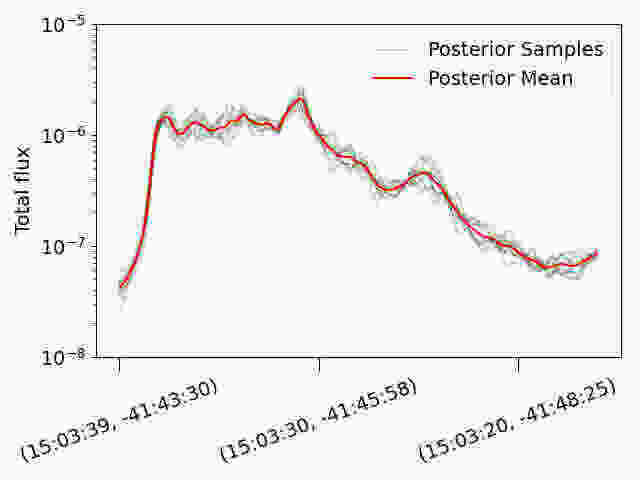}
  \includegraphics[width=0.3\textwidth]{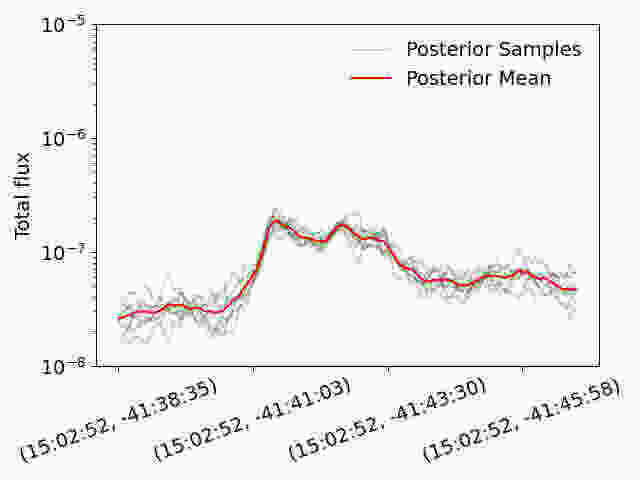}
  \includegraphics[width=0.3\textwidth]{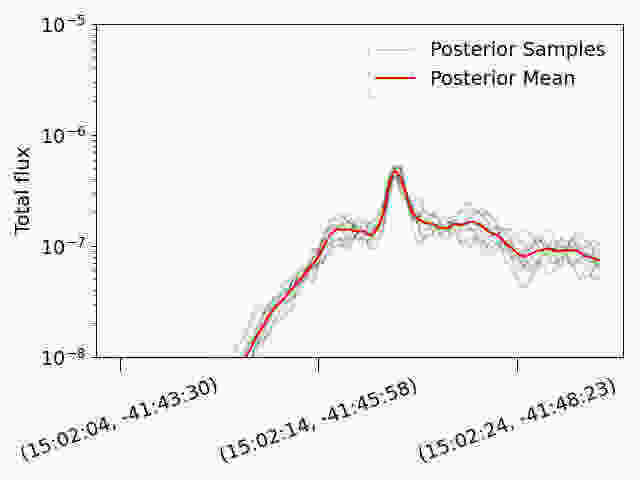}
  
  \includegraphics[width=0.3\textwidth]{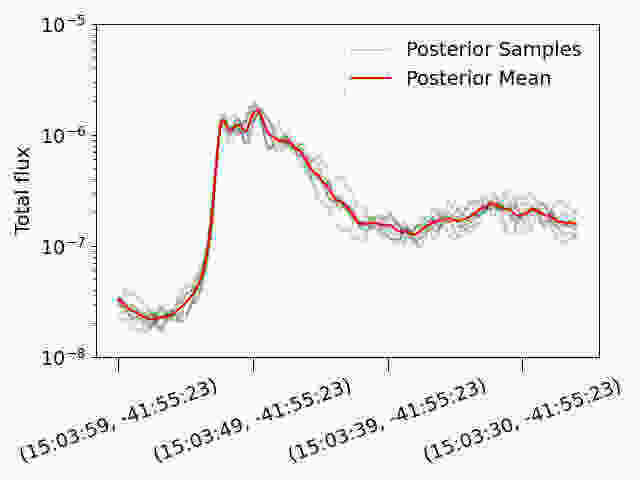}
  \includegraphics[width=0.3\textwidth]{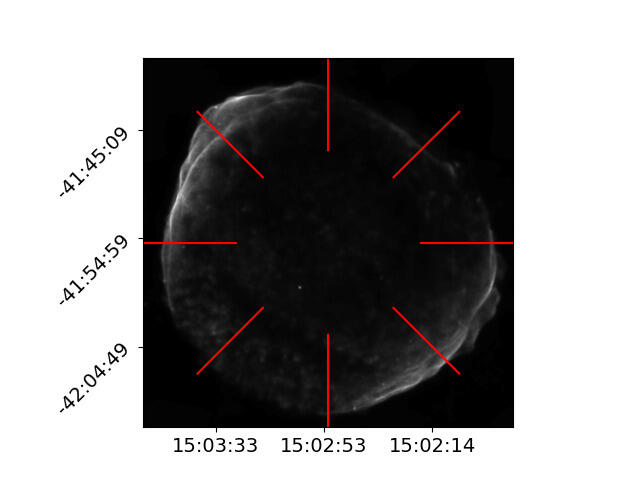}
  \includegraphics[width=0.3\textwidth]{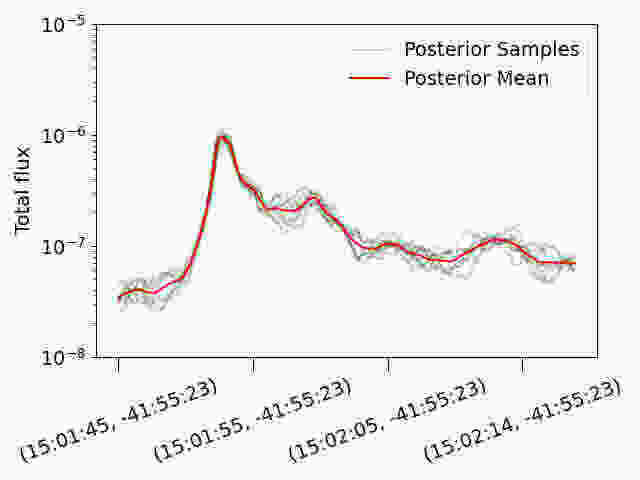}
  
  \includegraphics[width=0.3\textwidth]{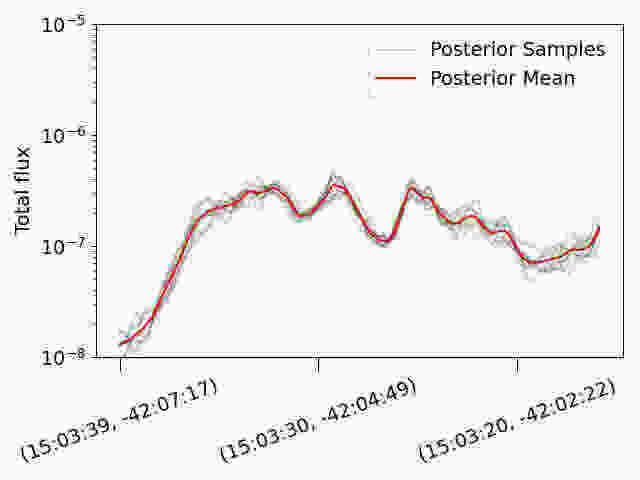}
  \includegraphics[width=0.3\textwidth]{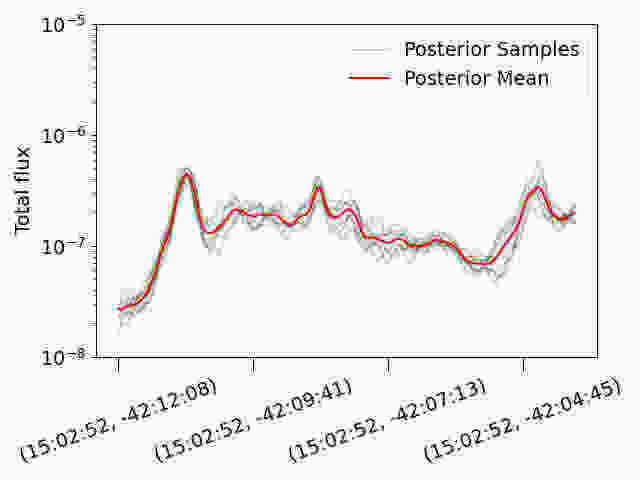}
  \includegraphics[width=0.3\textwidth]{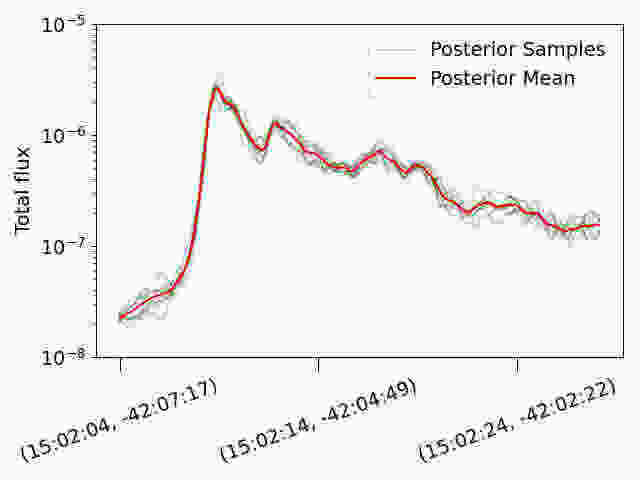}
  \caption{Flux intensity profiles in $[\text{s}^{-1}~\text{cm}^{-2}]$: The center image shows the location of the lines along which we present the intensity profile in pixel coordinates. The corresponding intensity profiles are plotted next to the line. The posterior mean of the reconstructed flux is plotted in red and the corresponding posterior samples are plotted in grey. The profiles are shown from left to right from the outsides to the insides of the \ac{SNR}.}
  \label{fig:intensity_profiles}
\end{figure*}
\subsection{Quantitative discussion}
\label{sec:QuantitativeDiscussion}
Fig. 8 shows the final results of applying our reconstruction algorithm to SN1006: the reconstructed sky and its separated components. The overall separation of diffuse emission and point sources succeeded. The point sources are clearly identified, and the \ac{PSF} deconvolution is particularly evident for the point sources. The diffuse structures are almost free of point source contributions. The effect of component separation can be seen more clearly in Fig. \ref{fig:Remnantzoom}, which shows a zoom on the north-eastern quarter of the remnant and its components. It can also be seen that some of the additional background noise from the \ac{BI} chip is partially absorbed into the background model.\\
One soft X-ray point source in the center of the remnant was not well separated from the diffuse emission. We believe this was caused by \ac{PSF} mismodeling in the outer pixels of the detectors, where the source is located in all observations considered, due to the assumption of invariant \ac{PSF}s within each observation. Fig. \ref{fig:psf_exposure} shows the exposure maps of the observations that covered this source and the position of the point source within these exposure maps. The pointing of ACIS-I and ACIS-S described in \citet{ChandraPOG} suggests large deviations of the actual PSF from the PSF model used in the positions of the mismodeled point source in detector coordinates. Dealing with position dependent \ac{PSF}s will be addressed in a future publication.\\
Fig. \ref{fig:smallzooms} shows the reconstruction of diffuse emissions from the remnant in detail. In order to study the remnant effectively, it is crucial to get a detailed view of it. To improve the clarity of our results, we present four close-up images of the remnant, highlighting its small-scale structures. The analysis shows that the shell is denoised both in the NW region, where we expect thermal emission, and in the SW region, where we expect non-thermal emission. We can also see that the denoising has improved the resolution of the small-scale structures in the inner X-ray emission of the remnant with respect to the data and also in comparison to the previous study of \citet{winkler2014high}.\\
Due to the statistical approach presented in this study, we get an estimate of the sky flux via the mean of the posterior probability, but also an uncertainty estimate via its standard deviation. The corresponding standard deviations for each energy bin are shown in the appendix in Fig. \ref{fig:uncertainty}. The top row of the figure shows the different energy bins of the posterior mean for better comparison. It can be seen that, as expected, the standard deviation is higher for regions of higher flux.\\
Thanks to the probabilistic approach we've adopted, we gain the ability to draw posterior samples from the inferred distribution. Such posterior samples $s^*$ allow us to compute the posterior mean, the standard deviation or any other statistical quantity of interest. Correspondingly, we can calculate the absolute \ac{NWR} $(\epsilon_\text{NWR})_j$ for each data patch $j$ as another interesting analysis of our results,
\begin{align}
\label{eq:nwr}
(\epsilon_\text{NWR})_j = \bigg \langle \frac{|d_j - \lambda_j(s)|}{\sqrt{\lambda_j(s)}} \bigg \rangle_{\mathcal{Q}(s|d)} 
\approx \frac{1}{N} \sum_{i=1}^N \frac{|d_j-\lambda_j(s_i^*)|}{\sqrt{\lambda_j(s_i^*)}}.
\end{align}
Here, $\lambda_j$ describes the reconstructed expected number of counts for each pixel in the data patch $j$. The absolute \ac{NWR}s provide a way to quantify the difference between a measured data point and its reconstruction, and help to distinguish between true deviations of the data from the reconstruction and deviations that are simply due to Poisson noise. The plots of the absolute \ac{NWR}s are shown in the appendix in Table \ref{tab:nwr} for each energy bin and data patch. These plots can be used as a sanity check on the correctness of the reconstruction presented here, as they allow to point out systematically unmodeled effects in the likelihood and the prior. We can see that the \ac{NWR}s are close to one in most regions, which implies a well-fitting model and reconstruction. In particular, in regions around point sources or at strong edges, we find higher \ac{NWR}s, which we attribute to the well-functioning deconvolution in these regions, leading to deviations of the reconstructed signal from the data.
\subsection{Analysis and comparison with previous studies}
As mentioned above, SN1006 has been studied extensively using a variety of instruments. In particular, several studies using X-ray telescopes have produced images of the \ac{SNR}. These studies have provided important insights into its structure and evolution, thus advancing our understanding of supernova explosions and their aftermath. Corresponding imaging approaches to SN1006 can be found in \citet{Winkler_2003}, \citet{10.1093/mnras/stv1882}, \citet{2003ApJ...589..827B} and also in \citet{CIAO}, which demonstrated the fidelity of the Chandra data processing pipeline for SN1006. In particular, in this study we have focused on the data and energy regions used by \citet{winkler2014high} and compare our reconstructions with their results. In \citet{winkler2014high} a comparison is made between the X-ray image and a H$\alpha$ image of SN1006 from CTIO. The comparison shows that there are several thin arcs of Balmer emission in the southern part of the remnant, which lie just in front of small-scale structures in the X-ray emission. In Fig. \ref{fig:smallzooms} we show these central parts of the remnant, which are dominated by soft X-rays. We show the enlarged cut-outs of these regions for the extracted remnant. Compared to previous studies, small-scale structures have an improved resolution due to the denoising and deconvolution, and are well disentangled from any point sources in the background. Accordingly, the presented reconstruction provides a more detailed view of the inner part of the remnant, enabling for a more accurate study of its small-scale structures.\\  
As noted in \citet{10.1093/mnras/stv1882}, the different energy bands show spatial variations in the remnant SN1006 with respect to each other. Fig. \ref{fig:reconstruction_result} shows these differing spatial variations of intensity with different X-ray energy bands. In particular, in Fig. \ref{fig:Remnantzoom} and Fig. \ref{fig:smallzooms} the parts of the hard X-ray lobe, the non-thermal regions in the \ac{SW} and \ac{NE} of the remnant, are well resolved and denoised, without any point source contribution. Soft X-rays in the NW shell are shown in Fig. \ref{fig:smallzooms}, which shows the shell of the thermal emission. \citet{koyama_evidence_1995} presented the first observational evidence that supernova shocks produce cosmic rays. However, the details of the acceleration mechanism of the particles is still an open question. Therefore, the study of the shock fronts is important to gain further insight into the acceleration mechanism and the dynamics of the shock front. The separation of the diffuse emission from the remnant allows us to visualize long intensity profiles along the remnant. Fig. \ref{fig:intensity_profiles} shows such radial intensity profiles of the \ac{SNR} in eight equidistantly space orientations. These denoised and deconvolved profile lines can be very useful to search for halos in front of the non-thermal regions and to get insights into the magnetic field strength according to \citet{Helder_2012}. Here, the profile lines show strong and sharp X-ray flux increases by up to two orders of magnitude at the shock front in the non-thermal regions in the \ac{SW} and \ac{NE} of SN1006. Notwithstanding, we defer the analysis of the structure of the remnant based on our reconstructions to future work.
\section{Conclusion}
\label{sec:conclusion}
In summary, we present a technique for obtaining an estimate of the true sky photon flux from Chandra X-ray event data. By  the sky flux as a generative process, this method allows us to infer not only the flux itself, but also its correlation structure in its extended components. Based on \ac{IFT} and by usage of Bayes theorem, this method approximates the posterior probability of a signal given the data via geoVI in a problem-adapted latent space. This allows to draw posterior samples in order to compute the expected sky flux, the posterior uncertainty and further validation and diagnostic measures.\\
Modeling the different correlation structures of point sources, diffuse emission and background in the \ac{BI} chips, we are able to separate point-like, extended sources and the additional noise in the \ac{BI} chips from the sought-after signal. Compared to previous separation and source extraction techniques, which are usually specified to extract either point sources or extended sources, the inference based on \ac{IFT} accounts for both components jointly. In particular, we build a spatio-spectral model for the sky flux based on the D4PO algorithm implemented by \citet{platz2022multicomponent} and use it for the spatio-spectral reconstruction of the X-ray sky. Since the spatio-spectral reconstruction is computationally expensive for a large number of pixels, we introduce an accelerated inference model, called the transition model. In the transition model, we first perform a spatial reconstruction in a single energy band, which has almost one order of magnitude less degrees of freedom as the spatio-spectral reconstruction, making it computationally less complex. The result of the spatial reconstruction, which already contains a lot of information about the sky flux in an energy bin and about the component separation, is used as an initial condition for the spatio-spectral reconstruction.\\
A benefit of the here presented analysis is the ability to build mosaics of different observations via the sum of logarithmic likelihoods. Each likelihood has its own description of the instrument response. This approach solves the problem of modelling different \ac{PSF}s for the same source in different data patches. \\
We apply the spatio-spectral reconstruction to the latest Chandra observations of SN1006, presented by \citet{winkler2014high}. The resulting image is a denoised, deconvolved and decomposed image, which provides a detailed view of the small-scale structures of SN1006. We reconstruct a separate image of the point sources present in the considered datasets, which can be compared with point source catalogs and, more importantly, allows us to study the X-ray emission of SN1006 without point source contributions. The different energy ranges in \ac{NE} and \ac{SW}, dominated by synchrotron emission, and the rest of the remnant, dominated by thermal emission, are clearly visible. The intensity profiles at the shell of the remnant are denoised and not visibly affected by point source contributions. We also show other diagnostics such as a simulated data reconstruction, uncertainties and noise-weighted residuals as a check for systematic errors. \\
Taking this work as a starting point for spatio-spectral Bayesian imaging of X-ray data, this study has pointed out the need for further methodological improvements. One is using a spatially varying \ac{PSF}. \citet{Alexander_2003} already showed the actual spatial variability of the \ac{PSF} in the Chandra image of the Deep Field North. We expect that the separation of the central point source in SN1006 to improve by the implementation of a spatially varying \ac{PSF}. This, however, is not a trivial task, as an invariant \ac{PSF} can be applied via multiplication in Fourier space, whereas a spatially varying \ac{PSF} cannot. Methods are currently being developed to solve this problem in the language of \ac{IFT}, including a neural network approach recently presented by \citet{e25040652}. In addition, a line model capable of modelling lines in the thermal emission will help to further resolve the energy direction. An interesting option here, already mentioned in \citet{seward_charles_2010}, would be to define different models for synchrotron emission and bremsstrahlung, to eventually be able to decompose the diffuse emission of the remnant into its thermal and non-thermal components. In general, we aim to further improve the reconstruction speed and reduce its computational cost to enable studies of more data sets, larger regions, and with higher resolutions in the spatial and spectral dimensions. As a contribution to this, we want to further optimize our hyper-parameter search to enable faster convergence of the algorithm.
\section{Acknowledgement}
M. Westerkamp, V. Eberle and M. Guardiani acknowledge support for this research through the project Universal Bayesian Imaging Kit (UBIK, Förderkennzeichen 50OO2103) funded
by the Deutsches Zentrum für Luft- und Raumfahrt e.V. (DLR). P. Frank acknowledges funding through the German Federal
Ministry of Education and Research for the project ErUM-IFT:
Informationsfeldtheorie für Experimente an Großforschungsanlagen
(Förderkennzeichen: 05D23EO1). J. Stadler and J. Knollmüller acknowledge funding by the Deutsche Forschungsgemeinschaft (DFG, German Research Foundation) under Germany's Excellence Strategy – EXC-2094 – 390783311.

\bibliographystyle{bibtex/aa}
\bibliography{ref}{}
\onecolumn
\appendix
\section{Synthetic data generation}
\label{sec:synthdata}
\begin{figure*}[!ht]
\centering
  \includegraphics[width=\textwidth]{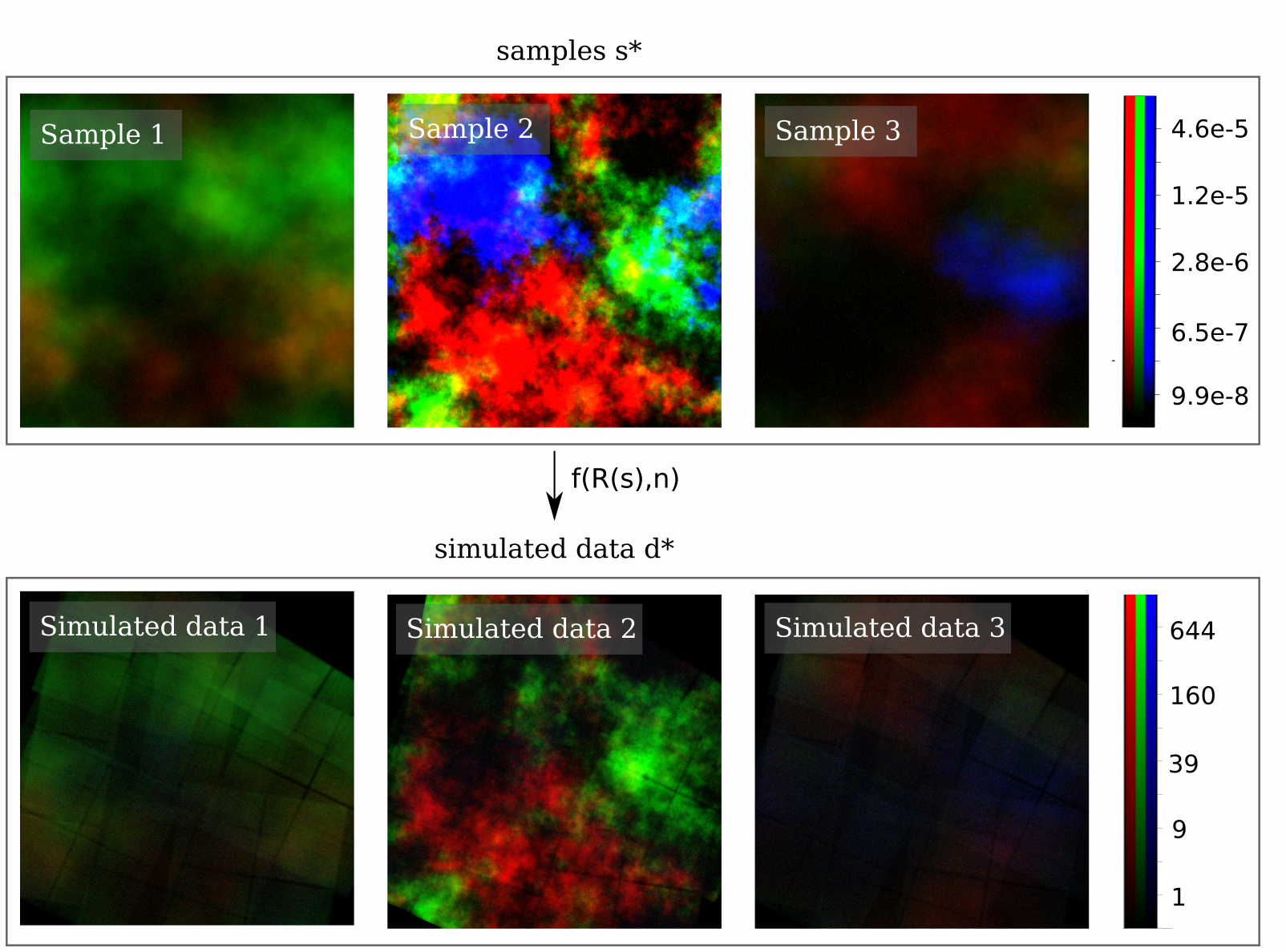}
\caption{Illustration of generation of simulated data for three prior samples, showing the variance in intensity and correlation structure permitted by the prior.}
\label{fig:priors}
\end{figure*}

Given the generative models, we can construct prior samples of the individual components and of the imaged sky composed of them, as described in Sect. \ref{sec:Priors}. Three of these prior samples are shown at the top of Fig. \ref{fig:priors}. They illustrate how the prior samples are converted into simulated data using the instrument response and mimicking Poisson noise. As mentioned in Sect. \ref{sec:Priors}, we use the prior samples and the simulated data to fine-tune the hyper-parameters prior to reconstruction. As we can see by comparison, the chosen hyper-parameters ensure that the order of magnitude of the data in Fig. \ref{fig:data} is the same as the order of magnitude of the simulated data. Moreover, and more importantly, the simulated data allow us to perform the validation of the algorithm as described in Sect. \ref{sec:mock}.

\newpage

\section{Computational analysis}
\label{sec:CompAnalysis}
\begin{figure*}[!h]
  \centering  
  \begin{subfigure}{0.48\textwidth}
   \includegraphics[width=\linewidth]{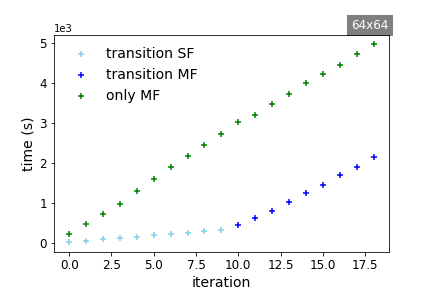}
   \end{subfigure}
  \begin{subfigure}{0.48\textwidth}
    \includegraphics[width=\linewidth]{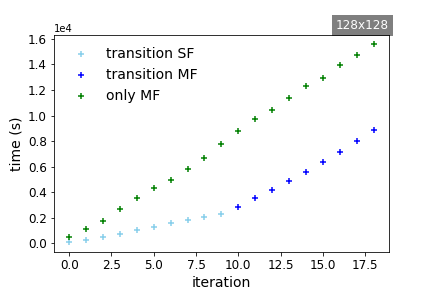}
    \end{subfigure} \\
   \begin{subfigure}{0.48\textwidth}
    \includegraphics[width=\linewidth]{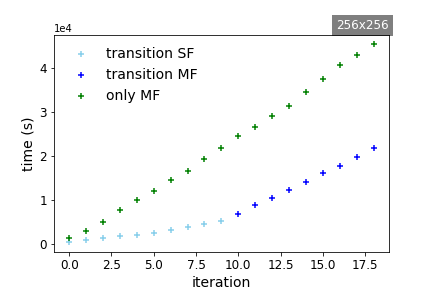}
    \end{subfigure}
       \begin{subfigure}{0.48\textwidth}
    \includegraphics[width=\linewidth]{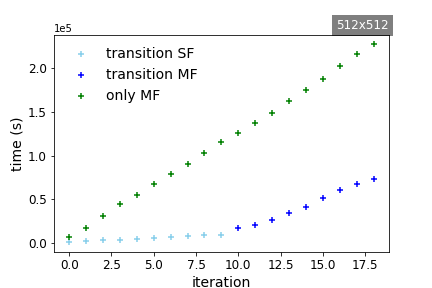}
    \end{subfigure} 
\caption{Time complexity (Left, top: $64 \times 64$ spatial pixels; Right, top: $128 \times 128$ spatial pixels; Left, bottom: $256 \times 256$ spatial pixels; Right, bottom: $512 \times 512$ spatial pixels). The time complexity is plotted for the different models. In green the only \ac{MF} reconstruction times per iteration are marked. In light blue the duration for each iteration in the \ac{SF} model before the transition is marked and in dark blue the duration of the \ac{MF} model iterations after the transition are shown. The first dark blue marker also includes the transition time.}
    \label{fig:comptime}
\end{figure*}

\begin{figure*}[!h]
  \centering  
      \begin{subfigure}{0.33\textwidth}
   \includegraphics[width=\linewidth]{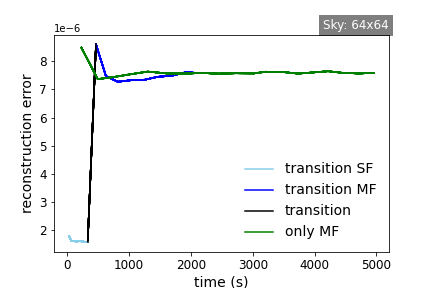}
   \end{subfigure}
          \begin{subfigure}{0.33\textwidth}
   \includegraphics[width=\linewidth]{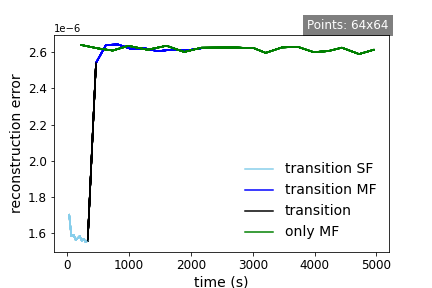}
   \end{subfigure}   
   \begin{subfigure}{0.33\textwidth}
   \includegraphics[width=\linewidth]{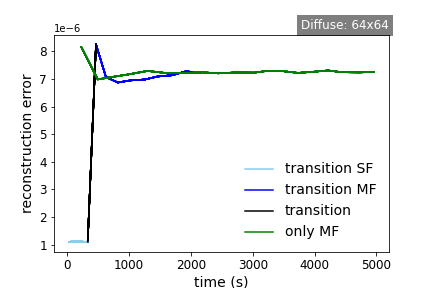}
   \end{subfigure} \\

  \begin{subfigure}{0.33\textwidth}
    \includegraphics[width=\linewidth]{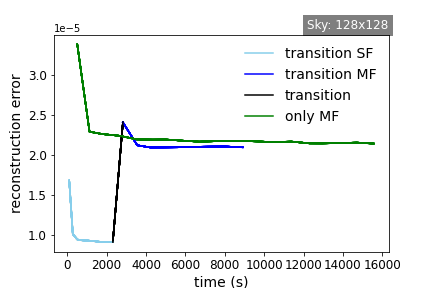}
    \end{subfigure}
  \begin{subfigure}{0.33\textwidth}
    \includegraphics[width=\linewidth]{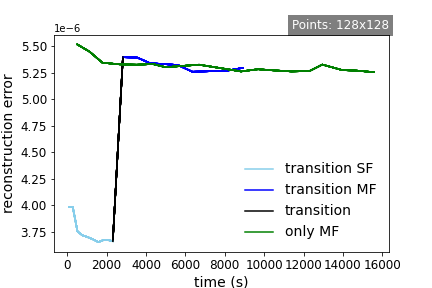}
    \end{subfigure}
      \begin{subfigure}{0.33\textwidth}
    \includegraphics[width=\linewidth]{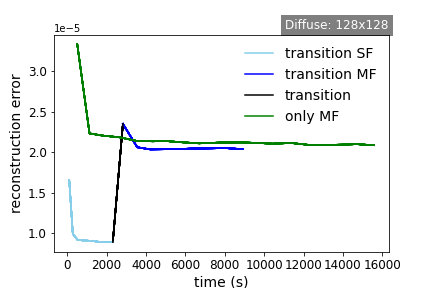}
    \end{subfigure} \\  
    
   \begin{subfigure}{0.33\textwidth}
    \includegraphics[width=\linewidth]{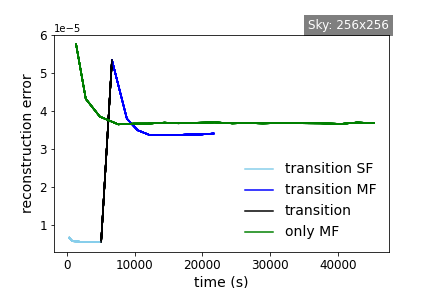}
    \end{subfigure}
       \begin{subfigure}{0.33\textwidth}
    \includegraphics[width=\linewidth]{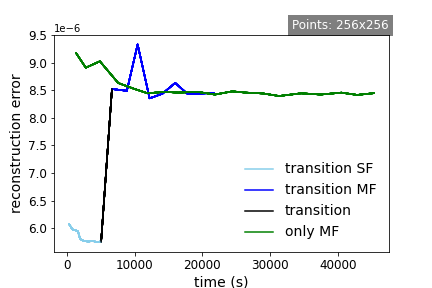}
    \end{subfigure}
       \begin{subfigure}{0.33\textwidth}
    \includegraphics[width=\linewidth]{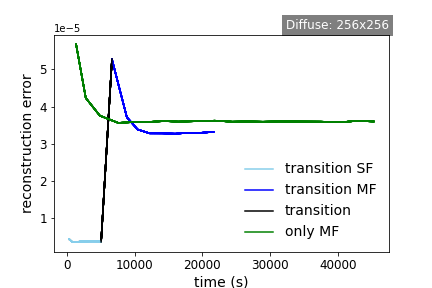}
    \end{subfigure} \\
    
       \begin{subfigure}{0.33\textwidth}
    \includegraphics[width=\linewidth]{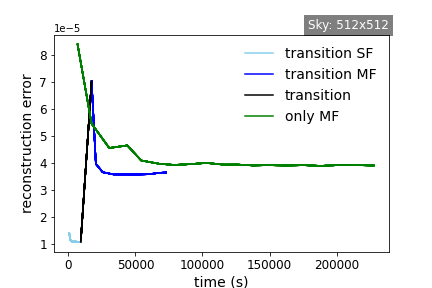}
    \end{subfigure} 
       \begin{subfigure}{0.33\textwidth}
    \includegraphics[width=\linewidth]{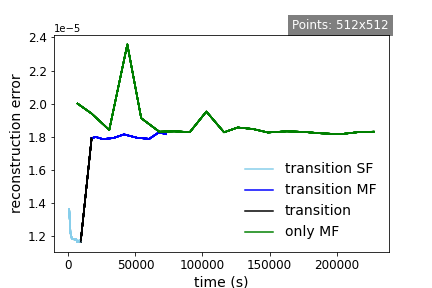}
    \end{subfigure} 
       \begin{subfigure}{0.33\textwidth}
    \includegraphics[width=\linewidth]{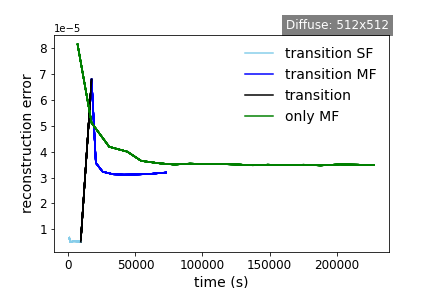}
    \end{subfigure} 
\caption{Reconstruction error in terms of the Frobenius norm (Eq.~\eqref{eq:FrobeniusNorm} from top to bottom for $64 \times 64$, $128 \times 128$, 
$256 \times 256$ and $512 \times 512$ spatial pixels and from left to right for the imaged sky, the point sources and the diffuse component. The green line marks the reconstruction error as a function of time for the pure \ac{MF} reconstruction. The light blue line marks the reconstruction error of the \ac{SF} reconstruction as part of the transition model, and correspondingly the black line marks the transition and the blue line marks the subsequent transition model \ac{MF} reconstruction.
In the iterations of the SF model, we typically anticipate lower reconstruction error in terms of small Frobenius norm. This expectation is attributed to the model's consideration of a smaller space.}
    \label{fig:compacc}
\end{figure*}

In this section, we present a comparison of the introduced algorithm including the transition model and a pure \ac{MF} reconstruction in terms of computational time and reconstruction error. In case of the transition model, we start with a \ac{SF} reconstruction and use the corresponding result as an initial condition for the \ac{MF} reconstruction, as described in Sect. \ref{sec:InferenceAlgorithm}. In the other case, we start the reconstruction on the whole \ac{MF} parameter space from the beginning. We consider four different spatial resolutions, from $64 \times 64$ to $512 \times 512$ pixels, for which we generate simulated data and perform the corresponding reconstruction on a single core for the transition model and the pure \ac{MF} model. This allows us to compare, for each problem size, the time complexity at each iteration and the reconstruction error as a function of time. \\
As already mentioned in Sect. \ref{sec:InferenceAlgorithm}, the parameter space for the \ac{SF} reconstruction is much smaller (Table \ref{tab:latent_parameter_table}), leading to higher computational time for each iteration in the \ac{MF} reconstruction. This effect is also shown in Fig. \ref{fig:comptime}. The time complexity of each iteration is accordingly higher in the \ac{MF}  reconstruction than in the \ac{SF} reconstruction, resulting in an overall lower time complexity for the transition model reconstruction. It can be seen that similarly to the duration of each iteration in the reconstruction, the transition time also increases with the growing parameter space. However, the increased transition time is not significant compared to the overall time savings.\\ 
More important for the analysis is how the reconstruction error of the reconstruction behaves over time. This is shown for the different components in Fig. \ref{fig:compacc}. As a measure of the reconstruction error, we compute the posterior sample mean for $N$ samples of the Frobenius norm of the sample residuals $r^*=(s^*-s_\text{gt})$ according to Eq.~\eqref{eq:Residuals} for each component,
\begin{align}
\langle \| r \|_F \rangle_{s^*} = \frac{1}{N}\sum_{n=0}^N	 \biggl(\sum_{i,j,k} (r^*_n)_{i,j,k}^2 \biggr) ^{\frac{1}{2}}, \label{eq:FrobeniusNorm}
\end{align}
where $i, j, k$ are the corresponding spatial and spectral pixel indices. Due to the consideration of a smaller space in the iterations of the SF model, we typically expect a smaller reconstruction error in terms of small Frobenius norm. It can be seen that the computational advantage of the transition model approach increases as the problem size increases in terms of a higher number of spatial pixels. This is especially true for the diffuse component, which is constructed from an outer product of correlated fields.

\newpage

\section{Further diagnostics for synthetic data reconstruction}
\label{sec:mock_diag}
\begin{figure*}[!h]
  \centering  
    \includegraphics[width=\textwidth]{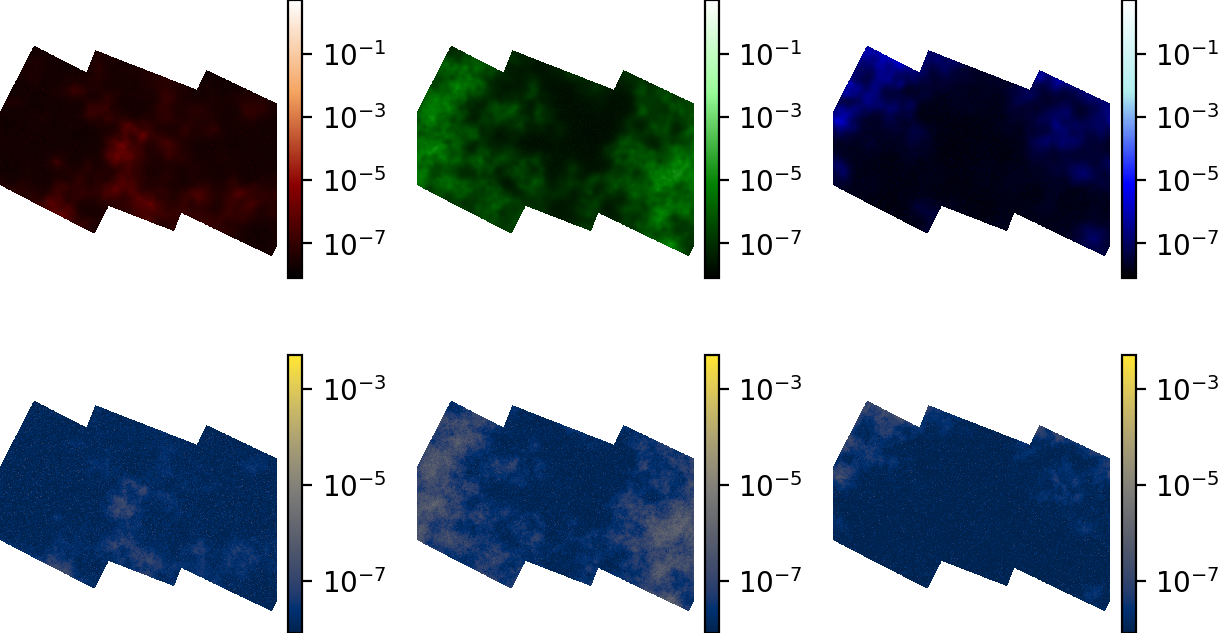}
    \caption{Synthetic data reconstruction uncertainties for the individual energy bins (left: 0.5-1.2 keV, center: 1.2-2.0 keV, right: 2.0-7.0 keV. Top row: Reconstruction results for the flux in $[\text{s}^{-1}~\text{cm}^{-2}]$ for the individual energy bins. Bottom row: Uncertainty maps for the individual energy bins.}
    \label{fig:mockskyuncertainty}
\end{figure*}

\begin{figure*}[!h]
  \centering  
  \begin{subfigure}{\textwidth}
  \centering
    \includegraphics[width=\textwidth]{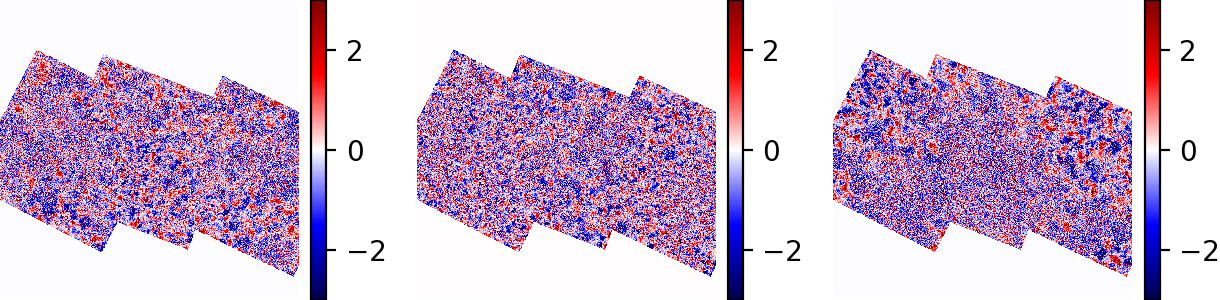}
    \end{subfigure}\\ \hfill
      \begin{subfigure}{\textwidth}
        \centering
    \includegraphics[width=\textwidth]{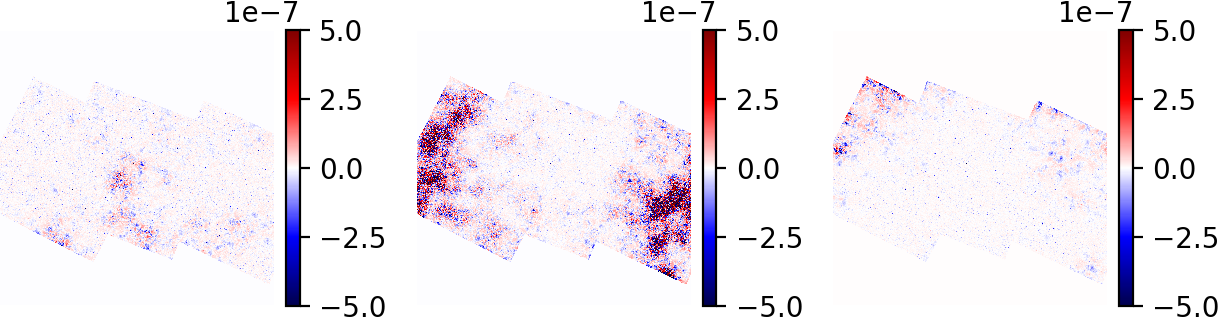}
    \end{subfigure}
\caption{Synthetic data reconstruction \ac{UWRs} (top row) and residuals (bottom row) for the individual energy bins (left: 0.5-1.2 keV, center: 1.2-2.0 keV, right: 2.0-7.0 keV) according to Eq.~\eqref{eq:uncertaintyweightedresiudal}.}
    \label{fig:mockskyuwr}
\end{figure*}

In this section, we present further diagnostic 
plots for sanity checks on the simulated data reconstruction in Sect. \ref{sec:mock}. The analysis of these plots can be found in the according sections. Fig. \ref{fig:mockskyuncertainty} shows the reconstruction results for the simulated data case for each energy bin together with the associated uncertainty. In addition, as a sanity check, we show the \ac{UWRs} and residuals for the simulated data case in Fig. \ref{fig:mockskyuwr}.


\newpage

\section{Further diagnostics for SN1006 reconstruction}
\label{sec:diag}
\begin{figure*}[!h]
  \centering
\begin{subfigure}{0.3\textwidth}
  \centering
  \includegraphics[width=\textwidth]{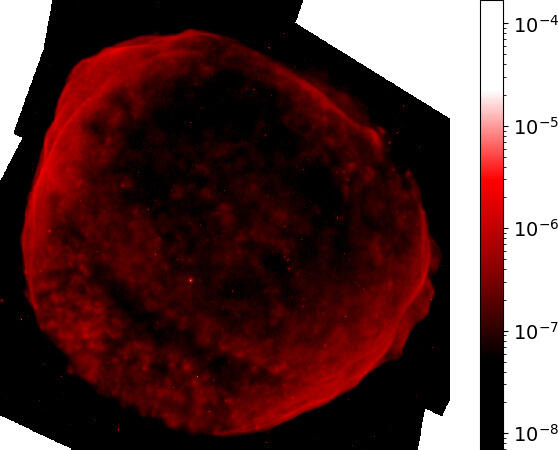}
\end{subfigure}
\begin{subfigure}{0.3\textwidth}
  \centering
  \includegraphics[width=\textwidth]{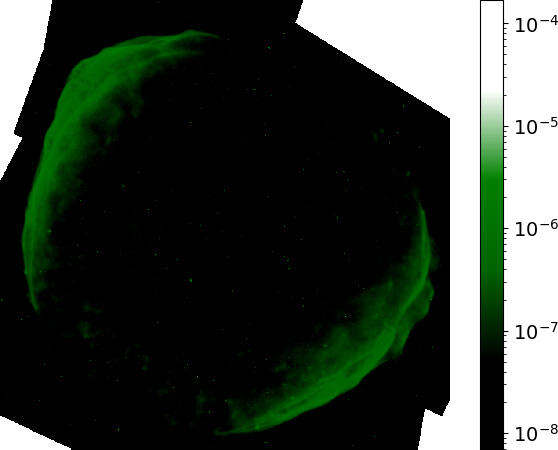}
\end{subfigure}
\begin{subfigure}{0.3\textwidth}
  \centering
  \includegraphics[width=\textwidth]{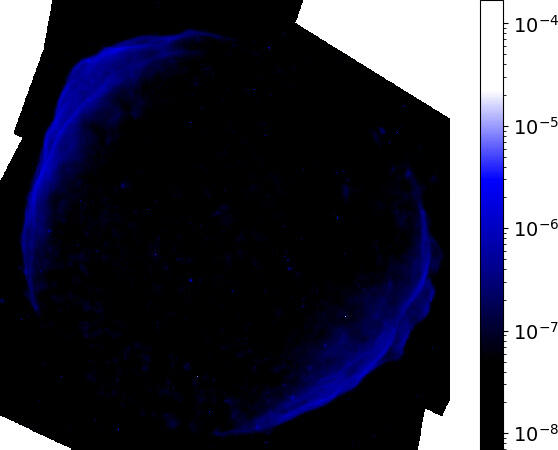}\end{subfigure} \\ 
  \centering
\begin{subfigure}{0.3\textwidth}
  \centering
  \includegraphics[width=\textwidth]{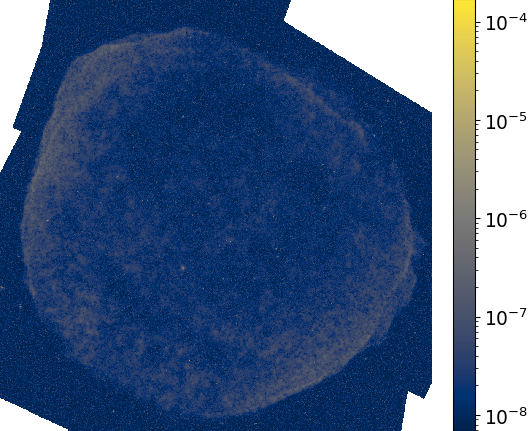}
  \end{subfigure}
\begin{subfigure}{0.3\textwidth}
  \centering
  \includegraphics[width=\textwidth]{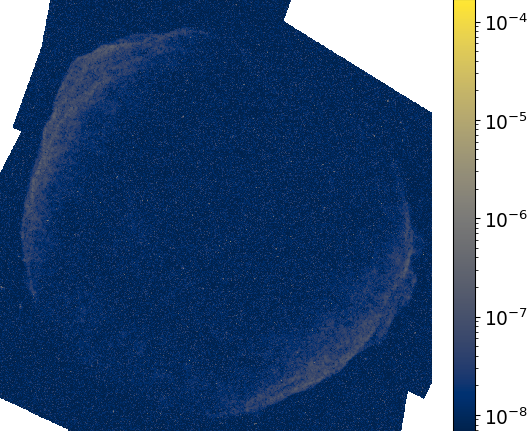}
  \end{subfigure}
\begin{subfigure}{0.3\textwidth}
  \centering
  \includegraphics[width=\textwidth]{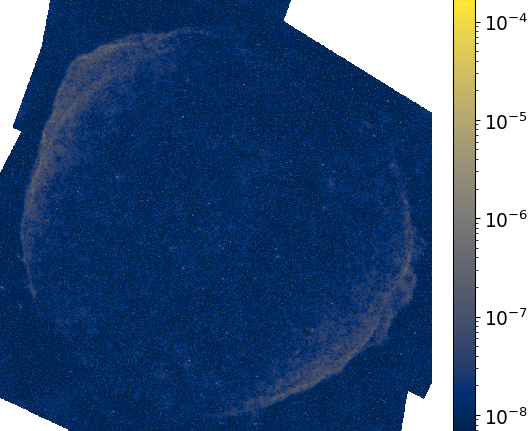}
    \end{subfigure}
  \caption{Posterior means and standard deviations for each energy bin in $[\text{s}^{-1}~\text{cm}^{-2}]$: Top row: Posterior means (red: 0.5-1.2 keV, green:1.2-2.0 keV, blue:2.0-7.0 keV). Bottom row: Posterior standard deviations (left: 0.5-1.2 keV, center:1.2-2.0 keV, right:2.0-7.0 keV).}
  \label{fig:uncertainty}
\end{figure*}

\begin{figure}[!h]
  \centering 
  \begin{subfigure}[t]{0.3\linewidth} 
    \includegraphics[width=\textwidth]{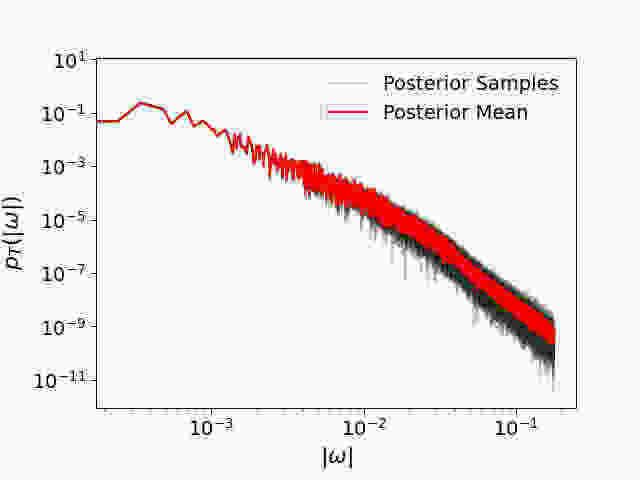}
 \end{subfigure}
   \begin{subfigure}[t]{0.3\linewidth} 
    \includegraphics[width=\textwidth]{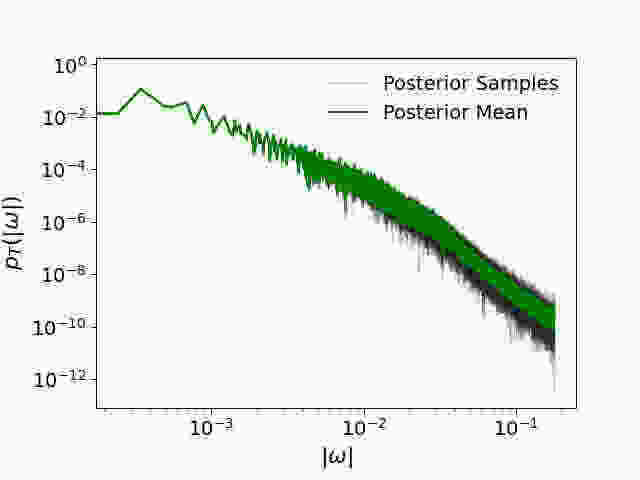}
 \end{subfigure}
   \begin{subfigure}[t]{0.3\linewidth} 
    \includegraphics[width=\textwidth]{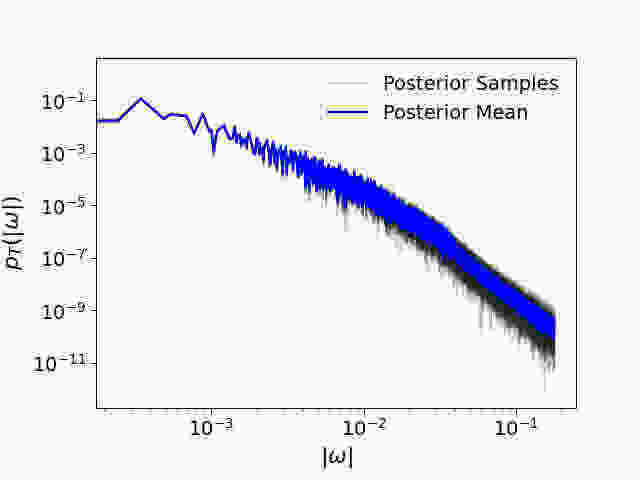}
 \end{subfigure}
 \caption{Spatial power spectra of the posterior mean and samples in the diffuse component for each energy bin (left: 0.5-1.2 keV, center: 1.2-2.0 keV, right: 2.0-7.0 keV)}
 \label{fig:spatial_pspec}
\end{figure}

Here, we show more diagnostic plots for the analysis of the reconstruction results presented in Sect. \ref{sec:results}. For further analysis of the reconstruction of the sky flux of the remnant SN1006, we present the posterior standard deviation separately for each energy bin and accompanied by the corresponding color bars in Fig. \ref{fig:uncertainty}. Furthermore, the reconstruction mean and posterior samples of the spatial power spectrum are shown in Fig. \ref{fig:spatial_pspec}. Finally, Table \ref{tab:nwr} shows the \ac{NWR}s according to Eq.~\eqref{eq:nwr} for each energy bin and dataset.

\begin{longtable}{|c|c|c|c|c|c|}
    \hline
    \textbf{Patch ID} & \textbf{0.5-1.2keV} & \textbf{1.2-2.0keV} & \textbf{2.0-7.0keV} \\
    \hline
    9107 & \includegraphics[width=0.25\linewidth]{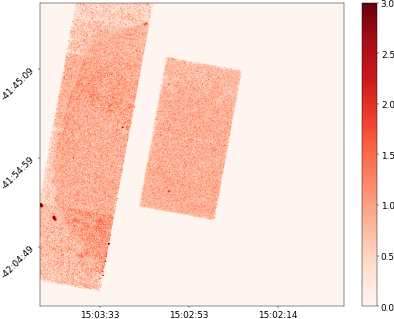}  & \includegraphics[width=0.25\linewidth]{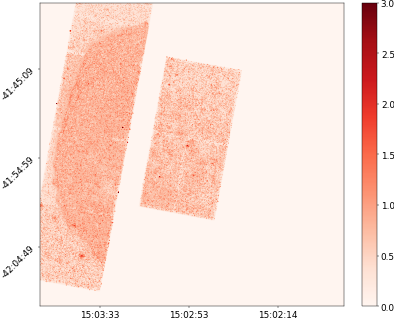} & \includegraphics[width=0.25\linewidth]{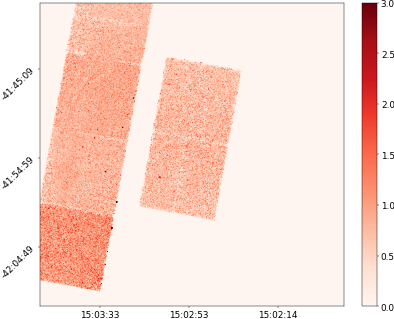}  \\
    \hline
        13737 & \includegraphics[width=0.25\linewidth]{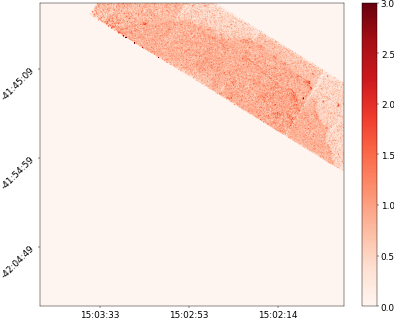}  & \includegraphics[width=0.25\linewidth]{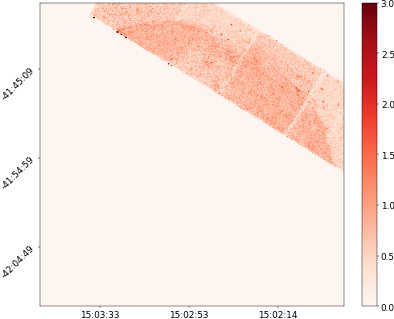} & \includegraphics[width=0.25\linewidth]{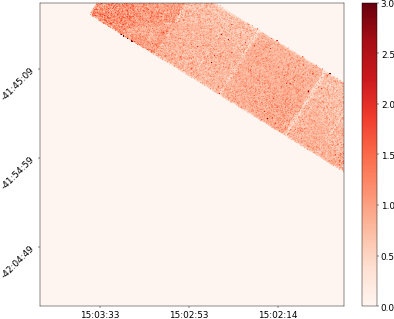}  \\
    \hline
            13738 & \includegraphics[width=0.25\linewidth]{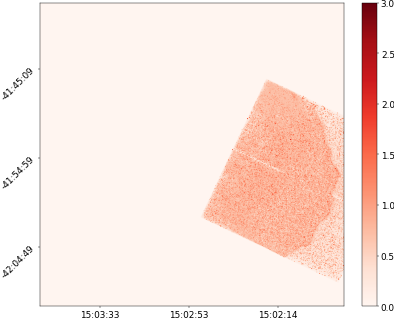}  & \includegraphics[width=0.25\linewidth]{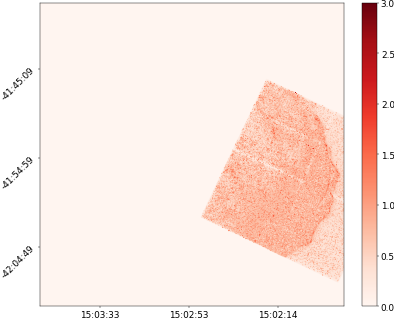} & \includegraphics[width=0.25\linewidth]{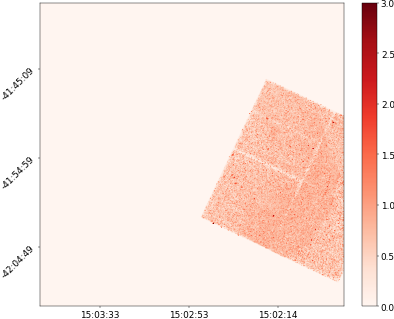}  \\
    \hline
                13739 & \includegraphics[width=0.25\linewidth]{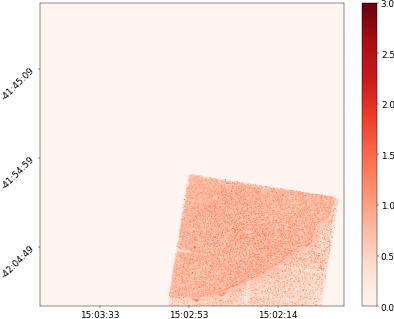}  & \includegraphics[width=0.25\linewidth]{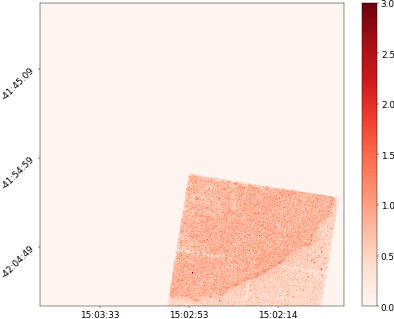} & \includegraphics[width=0.25\linewidth]{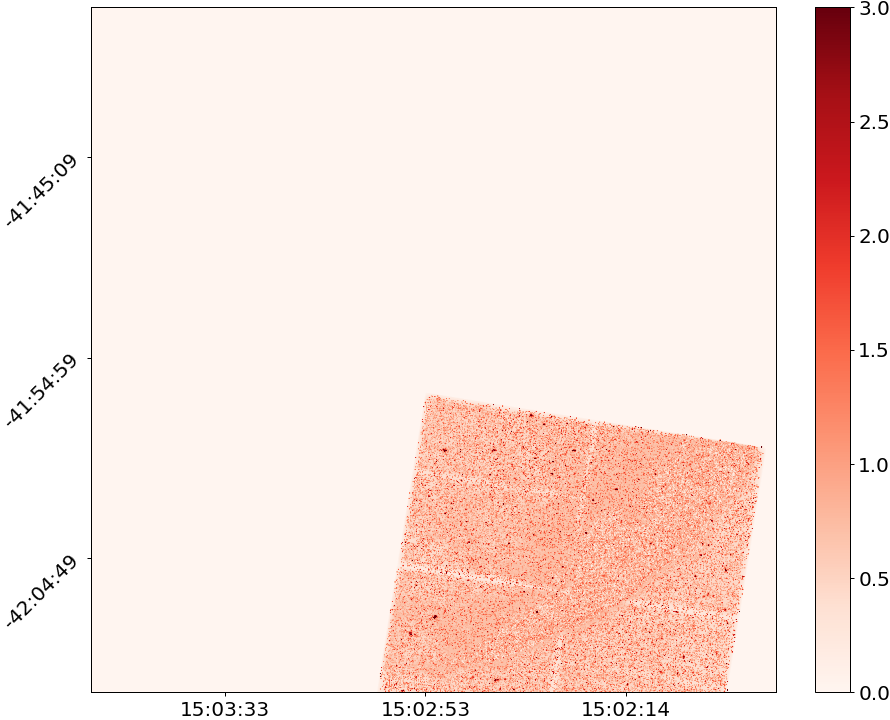}  \\
    \hline
                    13739 & \includegraphics[width=0.25\linewidth]{figs/13739sky_13739_nwr_0.png}  & \includegraphics[width=0.25\linewidth]{figs/13739sky_13739_nwr_1.png} & \includegraphics[width=0.25\linewidth]{figs/13739sky_13739_nwr_2.png}  \\
    \hline
                        13740 & \includegraphics[width=0.25\linewidth]{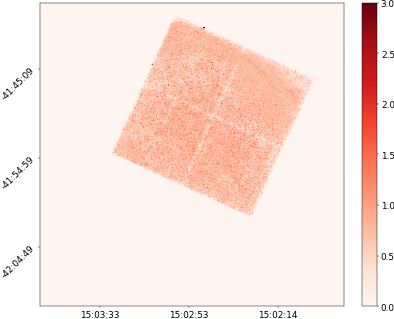}  & \includegraphics[width=0.25\linewidth]{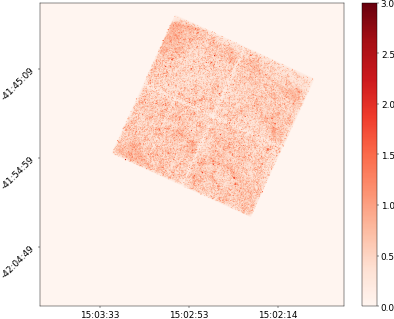} & \includegraphics[width=0.25\linewidth]{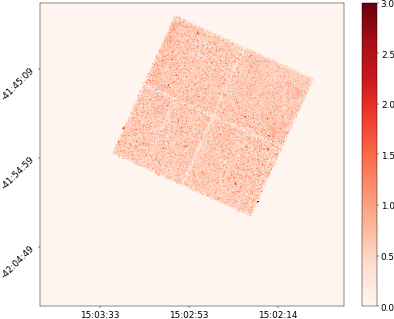}  \\
                        \hline
                        13741 & \includegraphics[width=0.25\linewidth]{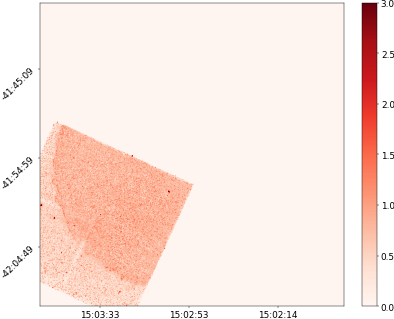}  & \includegraphics[width=0.25\linewidth]{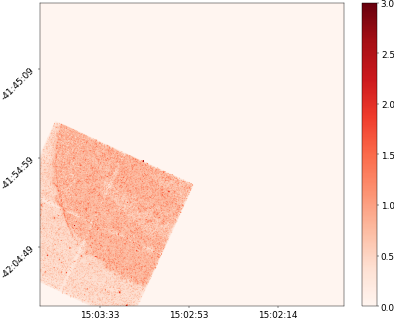} & \includegraphics[width=0.25\linewidth]{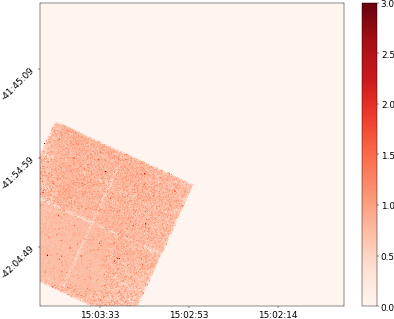}  \\
                        \hline             
                         13742 &
\includegraphics[width=0.25\linewidth]{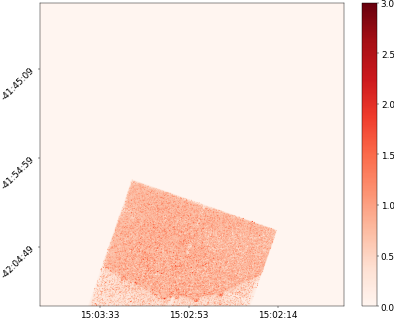}  & \includegraphics[width=0.25\linewidth]{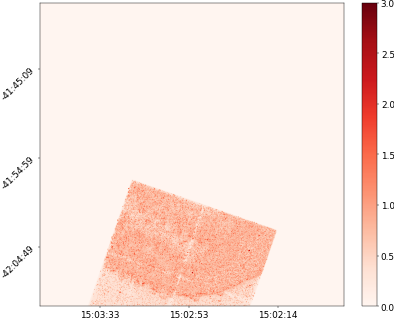} & \includegraphics[width=0.25\linewidth]{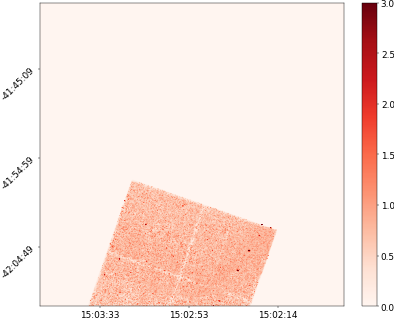}  \\
                        \hline             13743 &
\includegraphics[width=0.25\linewidth]{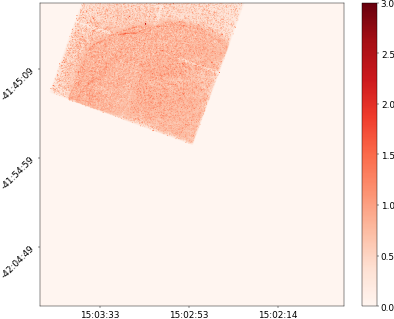}  & \includegraphics[width=0.25\linewidth]{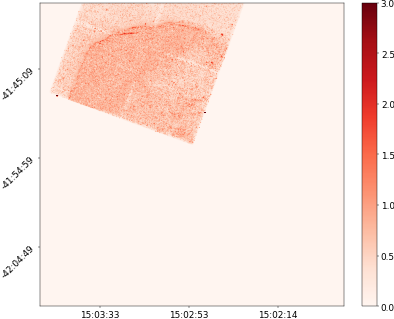} & \includegraphics[width=0.25\linewidth]{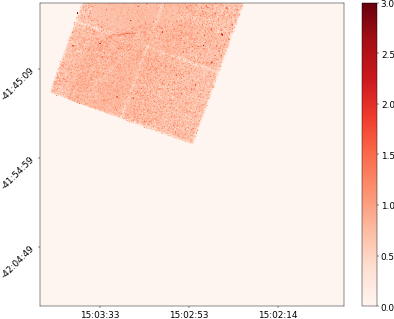}  \\
                        \hline  
                        14423 &
\includegraphics[width=0.25\linewidth]{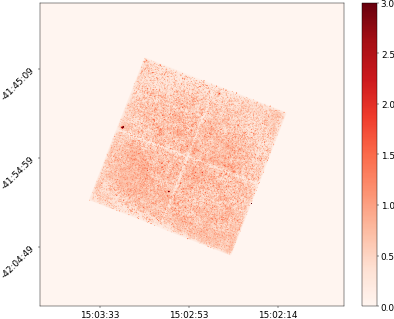}  & \includegraphics[width=0.25\linewidth]{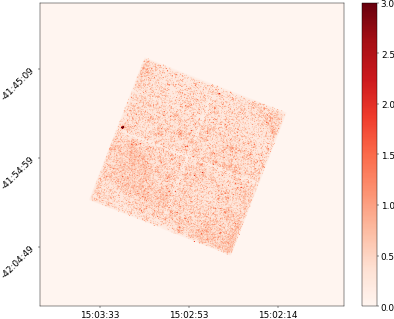} & \includegraphics[width=0.25\linewidth]{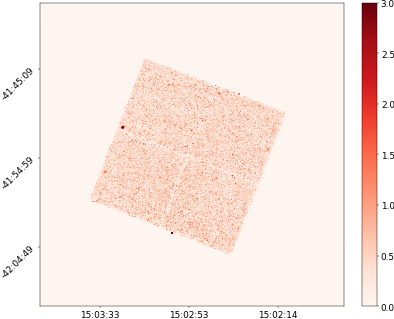}  \\
                        \hline   
                        14424 &
\includegraphics[width=0.25\linewidth]{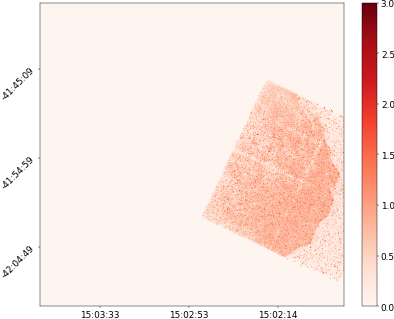}  & \includegraphics[width=0.25\linewidth]{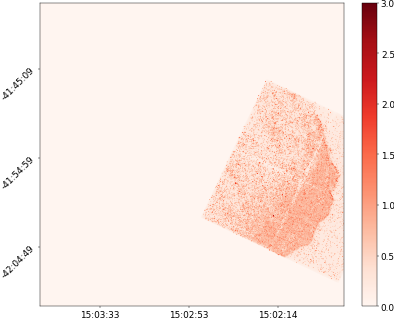} & \includegraphics[width=0.25\linewidth]{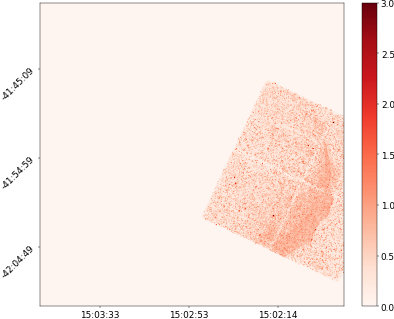}  \\
                        \hline  
                         14435 &
\includegraphics[width=0.25\linewidth]{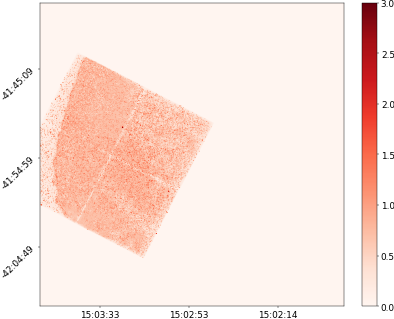}  & \includegraphics[width=0.25\linewidth]{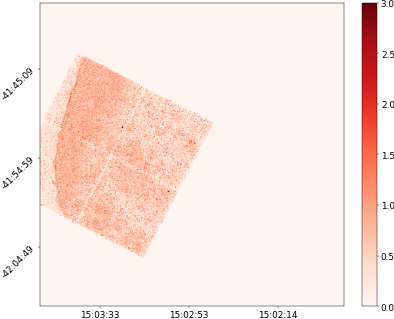} & \includegraphics[width=0.25\linewidth]{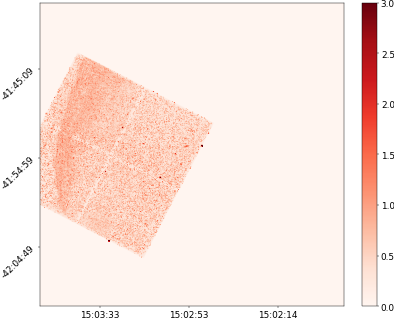}  \\
                        \hline                     
\caption{\ac{NWR} (Eq.~\eqref{eq:nwr}) for each dataset in Table \ref{tab:data_table} and energy bin.}
\label{tab:nwr}
\end{longtable}
\end{document}